\documentclass[acmsmall]{acmart}

\setcopyright{none}

\pdfoutput=1

\usepackage{amsmath} 
\usepackage{url}
\usepackage{booktabs}
\usepackage{multirow}
\usepackage{xspace}
\usepackage{mdframed}

\AtBeginDocument{%
  \providecommand\BibTeX{{%
    \normalfont B\kern-0.5em{\scshape i\kern-0.25em b}\kern-0.8em\TeX}}}

\newif\ifcomment

\ifcomment
  \newcommand{\gao}[1]{\textcolor[rgb]{0.3,0.8,0.4}{Gao Jie: #1}}
  \newcommand{\ken}[1]{\textcolor[rgb]{0.5,0.2,0.2}{Kenny: #1}}
  \newcommand{\simon}[1]{\textcolor[rgb]{0,0.8,0.4}{Simon: #1}}
  \newcommand{\added}[1]{\textcolor[rgb]{0.2,0.5,0.8}{#1}}
  \newcommand{\deleted}[1]{\textcolor[rgb]{0.8,0.8,0.8}{#1}}
\else
  \newcommand{\gao}[1]{}
  \newcommand{\ken}[1]{}
  \newcommand{\simon}[1]{}
  \newcommand{\added}[1]{\textcolor{black}{#1}}
  \newcommand{\deleted}[1]{}
\fi

\newcommand{\condA}{{\em{Condition A: Without AI, Asynchronous, not Shared Model}}\xspace}
\newcommand{\condB}{{\em{Condition B: With AI, Asynchronous, not Shared Model}}\xspace}
\newcommand{\condC}{{\em{Condition C: With AI, Asynchronous, Shared Model}}\xspace}
\newcommand{\condD}{{\em{Condition D: With AI, Synchronous, Shared Model}}\xspace}
\newcommand{\name}{{\emph{CoAIcoder}}\xspace}

\newcommand{\namef}{{\em CoAIcoder's\ }}

\newcommand{\ctime}{{\emph{Coding Time}}\xspace}
\newcommand{\irr}{{\emph{IRR}}\xspace}
\newcommand{\diver}{{\emph{Code Diversity}}\xspace}
\newcommand{\cover}{{\emph{Code Coverage}}\xspace}

\definecolor{figma_green}{HTML}{72c87a}
\definecolor{figma_blue}{HTML}{8fb5f9}
\definecolor{figma_orange}{HTML}{f18f6d}
\definecolor{figma_yellow}{HTML}{f5c242}

\setcopyright{acmcopyright}
\copyrightyear{2023}
\acmYear{2023}
\acmDOI{XXXXXXX.XXXXXXX}

\acmConference[Conference acronym 'XX]{Make sure to enter the correct
  conference title from your rights confirmation emai}{June 03--05,
  2018}{Woodstock, NY}
\acmPrice{15.00}
\acmISBN{978-1-4503-XXXX-X/18/06}

\begin{document}

\title[CoAIcoder: Examining the Effectiveness of AI-assisted \\ Human-to-Human Collaboration in Qualitative Analysis]{CoAIcoder: Examining the Effectiveness of AI-assisted Human-to-Human Collaboration in Qualitative Analysis}


\author{Jie Gao}

\affiliation{%
  \institution{Singapore University of Technology and Design}
  \country{Singapore}
}
\email{gaojie056@gmail.com}

\author{Kenny Tsu Wei CHOO}
\affiliation{%
  \institution{Singapore University of Technology and Design}
  \country{Singapore}}
\email{kenny@kennychoo.net}

\author{Junming Cao}
\affiliation{
  \institution{Fudan University}
  \city{Shanghai}
  \country{China}}
\email{21110240004@m.fudan.edu.cn}

\author{Roy Ka-Wei Lee}
\affiliation{
  \institution{Singapore University of Technology and Design}
  \country{Singapore}}
\email{roy_lee@sutd.edu.sg}

\author{Simon Perrault}
\affiliation{
  \institution{Singapore University of Technology and Design}
  \country{Singapore}}
\email{perrault.simon@gmail.com}

\renewcommand{\shortauthors}{Gao and Perrault et al.}



\begin{teaserfigure}
  \includegraphics[width=\textwidth]{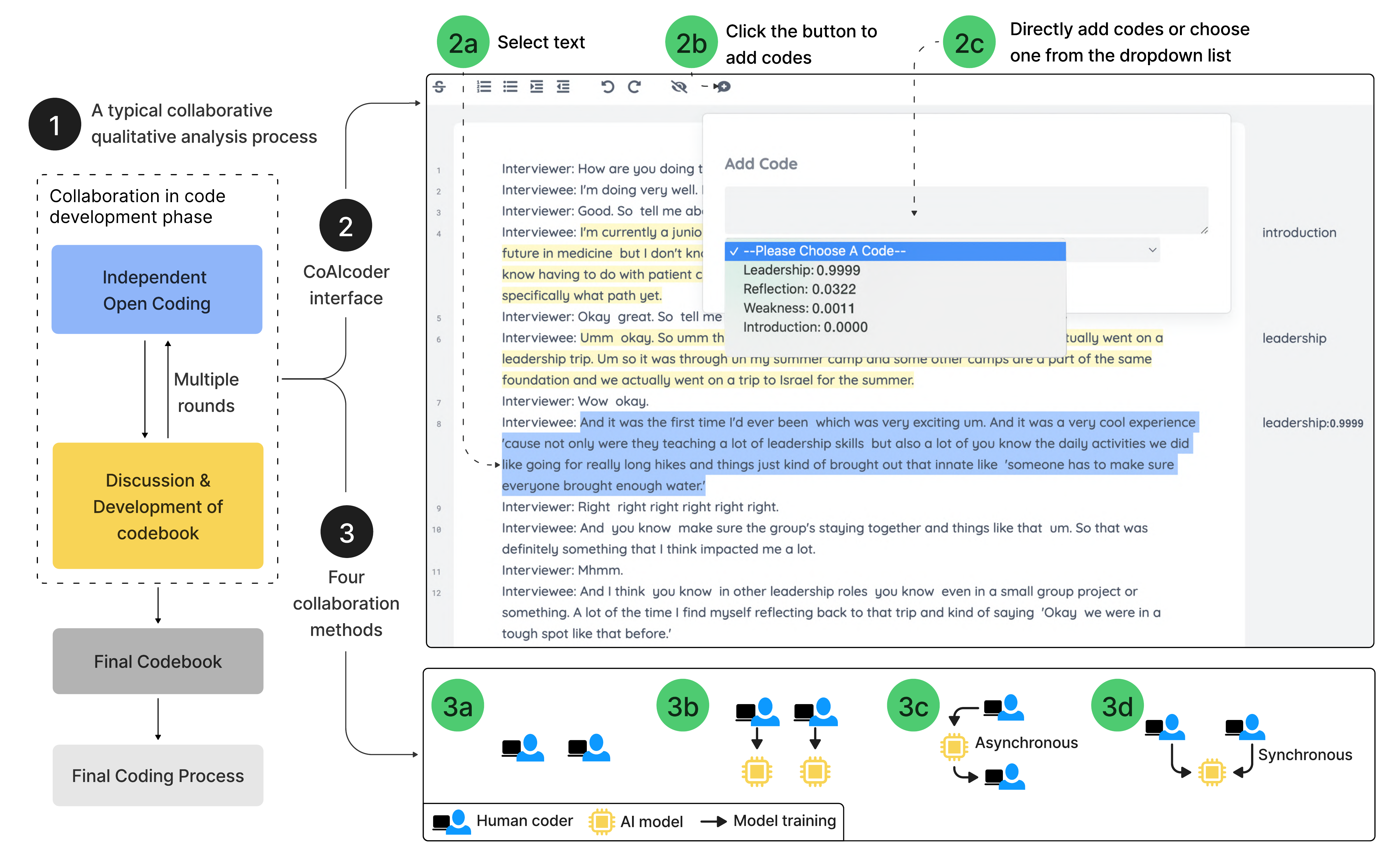}
  \caption{AI-assisted Collaborative Qualitative Analysis (CQA): 1) A typical CQA process, where collaboration mainly occurs during the coding development phase; 2) To explore the potential for AI to streamline this phase, we introduce \name, an \textbf{AI-assisted collaborative qualitative coding tool} that provides code suggestions with confidence levels based on users' coding history; 3) We also propose four distinct CQA methods to utilize \name: 3a) Without AI, Asynchronous, not Shared Model, 3b) With AI, Asynchronous, not Shared Model, 3c) With AI, Asynchronous, Shared Model, and 3d) With AI, Synchronous, Shared Model.}
  \label{fig:teaser}
\end{teaserfigure}

\received{06 April 2022}
\received[revised]{27 Nov 2022}
\received[accepted]{27 June 2023}


\begin{abstract}
While AI-assisted individual qualitative analysis has been substantially studied, AI-assisted \textbf{collaborative} qualitative analysis (CQA) – a process that involves multiple researchers working together to interpret data – remains relatively unexplored. 
After identifying CQA practices and design opportunities through formative interviews, we designed and implemented CoAIcoder, a tool leveraging AI to enhance human-to-human collaboration within CQA through four distinct collaboration methods.
With a between-subject design, we evaluated CoAIcoder with 32 pairs of CQA-trained participants across common CQA phases under each collaboration method.
Our findings suggest that while using a shared AI model as a mediator among coders could improve CQA efficiency and foster agreement more quickly in the early coding stage, it might affect the final code diversity.
We also emphasize the need to consider the independence level when using AI to assist human-to-human collaboration in various CQA scenarios.
Lastly, we suggest design implications for future AI-assisted CQA systems.
\end{abstract}

\maketitle

\section{Introduction}
Collaborative Qualitative Analysis (CQA) process enables the synthesis of diverse perspectives to establish collective interpretations within a team \cite{flick2013sage, richards2018practical}. This iterative procedure involves a thorough analysis of reading, identification, and coding of crucial data, followed by multiple rounds of discussions to form a codebook (see Figure \ref{fig:teaser}A) \cite{richards2018practical}. Despite its efficacy, the process is laborious and time-consuming, even for experienced researchers \cite{flick2013sage, marathe2018semi}.

Meanwhile, the emergence of Artificial Intelligence (AI), with its capacity to swiftly identify patterns in data, has been employed to enhance individual qualitative analysis (QA), thereby reducing the human effort required. For instance, QA is often interpreted as a classification task, utilizing algorithms such as Support Vector Machines (SVM) to categorize text into distinct code groups \cite{yan2014semi}. Additionally, QA can be seen as a topic modeling task \cite{bakharia2016interactive, gorro2017qualitative, lennon2021developing}, where the Latent Dirichlet Allocation (LDA) is employed to derive topics from the large-scale text, offering a comprehensive overview of the content.
Furthermore, QA can be approached as a keyword-matching task. This process involves using specific formulas (e.g., \texttt{definition OR define}) made from keywords to search for vector representations of text that contain these particular keywords. The search results are then associated with the corresponding codes that include the same keywords \cite{marathe2018semi, rietz2021cody}. 

\added{Despite the growing integration of AI in individual qualitative analysis, \textit{it remains unclear whether and how AI can be utilized to enhance the efficiency of CQA}, as the majority of current support tools for CQA largely depend on traditional techniques \cite{zade2018conceptualizing, ganji2018ease, drouhard2017aeonium}.} 
Therefore, we delve into this nascent area of \textbf{\textit{AI-assisted collaborative qualitative analysis}}, and utilize AI in the initial stages of CQA, where significant collaborations occur (see Figure \ref{fig:teaser}), with an aim to examine whether integrating AI has a potential to enhance the efficiency of human-to-human collaboration within the context of CQA. Our investigation distinguishes itself from existing methods that primarily tackle individual qualitative coding tasks, and more generally, human-AI collaboration \cite{rietz2021cody, marathe2018semi, kaufmann2020supporting, yan2014semi, bakharia2016interactive}.

To this end, we conducted a series of semi-structured interviews with QA researchers possessing different levels of experience. The findings from the interviews not only confirmed the primary steps involved in CQA~\cite{richards2018practical} but also shed light on our participants' expectations for the role of AI as a mediator among coders.

Based on these insights, we designed and implemented \name and its  AI \footnote{Natural Language Understanding (NLU) is an important part of Natural Language Processing (NLP), which falls under the broader umbrella of Artificial Intelligence (AI). In this work, for the sake of convenience and to maintain clarity, we will be using NLU, NLP, and AI interchangeably to refer to the language processing techniques and potential models employed.} model using the Rasa NLU pipeline\footnote{\url{https://rasa.com/docs/rasa/}} with a pre-trained word embedding. The model is fine-tuned in real-time, using the coding histories of the team as training data. Subsequently, when a user works on a new chunk of text, the AI model classifies the user's selected text into relevant classes and provides real-time code suggestions.
By doing this, we expect that users, upon encountering new codes from the suggested list, could promptly become aware of potential coding conflicts within their team. This would prompt them to actively scrutinize their coding decisions and foster mutual understanding among team members, resulting in a more refined discussion in later coding stages.

\added{Following that, to examine how AI can assist human-to-human collaboration within CQA, we designed four methods to use \name: \condA, \condB, \condC, and \condD (see Figure \ref{fig:teaser}(3)). 
The system, along with its four collaboration methods, was then evaluated using 32 pairs of participants, all well-trained in CQA. Each of the four methods was assigned to eight pairs.}

Our evaluation illuminates a key trade-off (i.e., coding efficiency vs. coding quality) and a crucial factor (i.e., the level of independence among coders) to consider when integrating an AI mediator into the initial stages of CQA: 1) low independence (i.e., more communication among coders) may achieve high efficiency and initial inter-rater reliability (IRR) but low code diversity; 2) high independence (i.e., less communication among coders) may achieve low efficiency and initial IRR, but higher code diversity.
Moreover, we highlight the importance of context in the application of AI-assisted CQA: whether a situation values efficiency or code diversity, and whether maximal coding efficiency is always beneficial. These aspects necessitate careful consideration, taking into account the specific requirements and context.

We contribute in the following ways:
\begin{itemize}
    \item Enhancing the understanding of CQA behaviors, challenges, and anticipated AI roles through semi-structured interviews. Identifying the potential of AI in facilitating human-to-human collaboration within the context of CQA. 
    \item Designing and implementing \name system, featuring four methods that augment human-to-human collaboration within CQA. The specific design features and guidelines introduced have the potential to inspire future research and development of AI-assisted CQA systems.
    \item Evaluating the four proposed AI-assisted CQA methods, demonstrating AI's potential to enhance the efficiency of human-to-human collaboration during the early stages of CQA, along with the associated trade-off.
\end{itemize}
\section{Related Work}
\subsection{Collaborative Qualitative Analysis}
Qualitative research is widely embraced across various disciplines, including but not limited to social science, anthropology, political science, psychology, educational research, and human-computer interaction \cite{charmaz2014constructing, corbin2014basics, lazar_research_2017}. CQA plays an important role in qualitative research and fosters robust and dependable interpretations of qualitative data \cite{anderson2016all, richards2018practical, flick2013sage}.
Cornish et al. provide a more detailed definition of CQA, describing it as \textit{"a process in which there is joint focus and dialogue among two or more researchers regarding a shared body of data, to produce an agreed interpretation"} \cite{cornish2013collaborative}.
Subsequently, Richards et al. delve deeper into the CQA methodology, breaking it down into a six-stage process \cite{richards2018practical}. The process starts with team planning, followed by open and axial coding. Then, the team formulates and tests an initial codebook. After passing this test, the final coding procedure occurs, concluding with a review of the codebook and the determination of overarching themes.

Although the CQA process has been broadly utilized \cite{corbin2014basics, flick2013sage}, its iterative and collaborative nature - encompassing reading, interpreting, discussing, and analyzing data - has been observed by researchers as a factor that contributes to its time-consuming nature \cite{ganji2018ease}.
In particular, the codebook development phase, including open coding and discussion, necessitates a substantial degree of collaboration.
Each coder independently adds codes and compares data, and the lead coder embarks on several iterative discussions with team members. The aim of these discussions is to categorize and structure the codes, leading to the creation of a codebook \cite{anderson2016all, richards2018practical, vollstedt2019introduction}.

\added{Recently, the emergence of AI has captured significant interest, particularly given its strong abilities for quickly understanding and making inferences from data~\cite{atlasAI, textminingIBM}.
In this work, we concentrate primarily on the integration of AI into the aforementioned iterative process. Our objective is to investigate whether employing AI could enhance the efficiency of discussion and communication among coders, a deviation from the traditional approach--communication among coders only commences following the open coding phase.}

\subsection{Approaches for Individual (Semi)Automatic Coding}
Due to limited research concerning the integration of AI into CQA, our review examines its utilization in individual QA. In particular, numerous solutions have been proposed to aid in individual qualitative coding. These include rule-based methods \cite{marathe2018semi, kaufmann2020supporting, rietz2021cody, crowston2010machine, gebreegziabher2023patat}, text classification \cite{rietz2021cody, yan2014semi, blackwell2018computer}, topic modeling \cite{leeson2019natural, bakharia2016interactive, leeson2019natural, lennon2021developing, baumer2017comparing}, and active learning approaches \cite{yan2014semi}, among others.

One significant approach involves rule-based methods, which utilize a combination of various keywords to match text \cite{marathe2018semi, kaufmann2020supporting, freitas2017learn}. For example, Marathe et al. \cite{marathe2018semi} presented a Python framework that uses coding rules, formulated manually and combined using Boolean operators such as AND, OR, and NOT. These are used to search for matching text within qualitative data and associate the text with codes when the cosine similarity exceeds a pre-set threshold.
\added{Rietz et al. \cite{rietz2021cody}, aiming to provide end-users with the flexibility to edit and refine coding rules during the coding process, introduced Cody, this interactive tool allows users to interactively modify rules generated by the model and define their own codes during the coding process.
These user-generated rules serve as the training data for a machine-learning model, leading to the better production of code suggestions.}

\added{Nevertheless, relying solely on code rules created by humans limits flexibility and necessitates additional human effort. As a result, researchers have proposed leveraging computer systems to detect code patterns \cite{nelson2020computational, gebreegziabher2023patat}. For instance, Nelson \cite{nelson2020computational} introduced a three-step approach (pattern detection, refinement, and confirmation) that utilizes techniques like unsupervised machine learning and word frequency scores. These techniques assist researchers in identifying novel patterns within their data, facilitating scalable and inductive exploratory analysis. To preserve users' control and agency in the qualitative coding process, Gebreegziabher et al. \cite{gebreegziabher2023patat} presented PaTAT, which identifies explainable patterns within the user's coding data simultaneously during the coding process, offering predictions for potential future codes. 
These works imply that the utilization of (semi)automatic code pattern detection for partial coding automation has significant promise.}

Furthermore, text classification techniques have gained widespread usage in qualitative analysis. For example, Yan et al. \cite{yan2014semi} proposed an approach that employed Support Vector Machine (SVM) classification, utilizing pre-selected features and parameters. The SVM model was trained using codes assigned by human coders, enabling the classification of large-scale text data.
Similarly, in Cody by Rietz et al. \cite{rietz2021cody}, a logistic regression model with stochastic gradient descent (SGD) learning was trained to classify unseen data based on the available annotations.

In addition, topic modeling techniques such as Latent Dirichlet Allocation (LDA) have been explored to unveil hidden topics within qualitative data \cite{baumer2017comparing, leeson2019natural, hong2022scholastic}. This is discovered to be similar to the outcomes of the results derived from open coding, indicating the potential application of topic modeling in qualitative analysis \cite{baumer2017comparing}.

\added{In conclusion, this review reveals potential for integrating AI into QA to enhance its efficiency. Similarly, we harnessed users' inputs as training data, to fine-tune the text classification model with pre-trained word embedding, which can provide code predictions promptly when users make requests during the coding process.}

\subsection{Existing Tools to Support CQA}
Numerous works have been investigated to assist CQA using traditional technologies.
For example, Zade et al. \cite{zade2018conceptualizing} proposed a strategy that enables coders to order the degrees of consensus, using tree-based ranking metrics to quantify coding ambiguity. This ordering can extend from the most ambiguous to the least, or from low to high agreement. Likewise, \textit{Aeonium}, introduced by Drouhard et al.\cite{drouhard2017aeonium}, is a visual analytics system that assists team members with tools to review, edit, and add code definitions and examples shared within the team. It also enables monitoring of the coding history throughout the process, with the aim of revealing disagreements and reducing ambiguity. 
Furthermore, Ganji et al. \cite{ganji2018ease} introduced \textit{Code Wizard}, a visualization tool embedded in Excel that leverages the certainty level of the codes assigned by all coders to highlight highly ambiguous codes. In addition, it enables the aggregation of individual coding tables, the automated sorting and comparison of coded data, and the calculation of inter-rater reliability. 
All the above works improve discussions between coders, allowing coders to focus on the most challenging aspects of the work. However, they caused the CQA process to diverge somewhat from the traditional coding procedure, introducing additional steps and consequently increasing the overall complexity~\cite{chinh2019ways}. As a consequence, these tools demand much learning and acclimatization, which could potentially pose new challenges for users.

On the other hand, current commercial CQA software like MaxQDA\footnote{\url{https://www.maxqda.com}}, Atlas.ti\footnote{\url{https://atlasti.com}}, nVivo\footnote{\url{https://lumivero.com/products/nvivo/nvivo-product-tour/}} and Google Docs~\cite{freitas2017learn, nielsen2018collaborative}, demonstrate a more intuitive and user-friendly approach, largely maintaining the familiarity of traditional coding procedures while offering the benefits of modern collaborative tools.
MaxQDA, for instance, provides a feature for merging coding documents from multiple coders once independent coding is complete \cite{MaxQDADocument, oswald2019improving, marathe2018semi}. The web version of Atlas.ti and Google Docs boasts even more advanced capabilities: they permit multiple users to code simultaneously and synchronize modifications in real time. The seamless integration of simultaneous coding and real-time synchronization significantly reduces workflow disruptions and promotes efficient and effective collaboration, addressing the issue of delayed information updates among team members, a common drawback found in other systems \cite{marathe2018semi}. However, we've noticed that tools like Atlas.ti and Google Docs allow users to code and view others' codes within a shared document. This functionality could potentially introduce substantial bias among coders \cite{anderson2016all}, as it means that a coder's work is continually visible to their peers. Therefore, we strive to strike a balance between collaborative efforts and individual perspectives through AI mediation, ultimately enhancing the quality and integrity of the coding process in CQA.

Meanwhile, there's a growing interest in leveraging AI to augment CQA. For example, Rietz et al.~\cite{rietz2021cody} recognized the potential of AI in CQA and call for efforts to examine the extent to which code rules and formulas can assist multiple coders in discussing their interpretations of codes during the coding process. The authors also suggested investigating how coders interact with suggestions generated based on their co-coder's coding rules. 
This underscores the prospective advantages that AI integration could introduce to the CQA process.

\added{In summary, while collaboration plays a crucial role in qualitative analysis, and AI has been extensively studied in individual qualitative analysis, the domain of \textbf{AI-assisted collaborative qualitative analysis} remains relatively unexplored. 
To fill this research gap, we conducted an investigation comprising a series of semi-structured interviews with HCI researchers possessing diverse levels of expertise.
Drawing on insights from our interviews, we developed \name, employing four distinct AI-assisted collaboration techniques in its evaluation. The focus of this evaluation was to investigate the impact of these collaboration techniques on the dynamics of coding efficiency and coding outcomes, in order to give us a deeper understanding of AI's potential in fostering human-to-human collaboration within the context of CQA.
Lastly, we seek to derive valuable design implications from our findings, which can serve as guidelines for developing more effective and efficient collaborative coding tools in the future.}

\section{Formative Interview}
\label{sec:interview}

To understand the current CQA practices, challenges, and users' anticipations when integrating AI, we conducted a series of semi-structured interviews with HCI researchers possessing varying levels of CQA experience.
Our university's Institutional Review Board (IRB) granted approval for these interviews.

\subsection{Methodology}

Our semi-structured interviews involved 8 HCI researchers (4 females, 4 males, mean age = 29.9 years), all of whom possessed experience in QA and CQA. The participants encompassed two postdoctoral researchers (P1, P2), two senior graduate researchers (P3, P4) regularly utilizing (C)QA in their work, and four junior graduate researchers (P5, P6, P7, P8) possessing a minimum of 1.5 years of experience in (C)QA. 
Further details, including their respective fields of study, educational background, software used for (C)QA, and the duration of QA use in their research, can be found in Table~\ref{table:userstudy}.

\begin{table}[!b]
\caption{Demographics of Interview Participants. Every participant was working in tje field of HCI and was a master's student or above.}
\label{table:userstudy}
\centering
\scalebox{0.78}{\begin{tabular}{|c|c|c|c|c|}
\hline
\multicolumn{1}{|c|}{\textbf{Participant ID}} & \multicolumn{1}{c|}{\textbf{\added{Fields of Study}}} &
\multicolumn{1}{c|}{\textbf{Education}} & \multicolumn{1}{c|}{\textbf{QA Software}} & \multicolumn{1}{c|}{\textbf{QA Experience (Years)}} \\ \hline
P1 & HCI, Mobile Computing & Postdoc & Excel & 9 \\ \hline
P2 & HCI, Healthcare & Postdoc & nVivo/Atlas.ti/MaxQDA/Excel & 7 \\ \hline
P3 & \added{HCI, Ubicomp} & Ph.D. student & Atlas.ti/Excel  & 4 \\ \hline
P4 & \added{HCI, AI} & Ph.D. student & Whiteboard/Google Sheet  & 4 \\ \hline
P5 & \added{HCI, VR} & Ph.D. student & Word/Excel/nVivo & 2 \\ \hline
P6 & HCI, Healthcare & Master's Student & Google Docs & 3.5 \\ \hline
P7 & \added{HCI, Chatbot} & Master's Student & Atlas.ti & 3 \\ \hline
P8 & HCI, Security, Privacy & Master's Student & Atlas.ti/Excel  & 1.5 \\ \hline
\end{tabular}}
\end{table}

During the interview, we asked participants to reflect upon and share their most notable CQA experiences. In addition to narrating their experiences, we requested participants to demonstrate their data coding process via screen sharing when feasible. We also explored the challenges they encountered with their chosen tools, as well as their expectations concerning potential AI assistance in the process. All interviews, lasting between 20 and 60 minutes, were conducted virtually, audio-recorded, and transcribed through Zoom to facilitate subsequent analysis.

In order to derive nuanced insights from our data, two authors, including the interviewer and an experienced qualitative researcher, employed a CQA process as per Richards et al. \cite{richards2018practical}. The coders began with independent open coding of two representative transcripts (P2, P3). Following this initial round, they convened to discuss similarities, reconcile code conflicts, and establish consensus on a primary codebook. The codebook was then tested and refined using two additional interview transcripts (P1, P4). Lastly, the first coder processed the remaining four transcripts (P5-P8). The findings from this process are presented in the following subsection.

\subsection{Findings}
\subsubsection{Basic CQA Process}
Our participants generally followed the main steps of the CQA process outlined by Richard et al. \cite{richards2018practical}. The following summarises their practices:

\begin{enumerate}
    \item Each coder in the team receives a copy of the data (e.g., interview transcripts, qualitative notes). They individually review the material and identify key points and primary codes through a process known as initial coding or open coding.
    \item The coders convene to discuss their respective codes and selected key points. During these discussions, they address any discrepancies or differences in understanding and interpretation of the codes. Through this, they aim to reach a consensus and propose a primary codebook. One participant described this process as follows:
    \textit{"The first level of analysis is like: I have a copy of the data, and my collaborator has another copy, and we'll assign primary codes. Then we sat down and discussed them, and see if he or she agrees. If there is disagreement, we will rethink the code and discuss." (P1, postdoc with 9 years of QA experience).}

    \item The coders proceed to code additional data and expand the content within the codebook. This iterative process is often repeated for several rounds until the codes stabilize. One participant explained this process as follows:
    \textit{"Based on the existing [codes in] open coding, we can roughly divide it into several pieces [or groups], and then pick out a few themes. [Sometimes] it is difficult to determine the theme, then there will be several rounds discussions [to decide the themes]." (P6, Master's student with 3.5 years of QA experience).}

    \item Once the codebook reaches a stable state, one or more coders employ this finalized codebook to code the remaining data. 

    \item The coders proceeded to generate reports based on their findings. These reports synthesized the coded data and identified emerging themes, patterns, and examples derived from the analysis. 
\end{enumerate}

However, in our inquiries about the specific methods employed for CQA, there were different responses regarding their preferred approaches. Some participants indicated utilizing grounded theory \cite{corbin2014basics, bryant2007sage}, while others mentioned adopting thematic analysis \cite{maguire2017doing, smith2015qualitative, braun2006using}. These methodological choices were influenced by the specific requirements and objectives of their respective projects. 

\added{Specifically, when time and resources permit, and when participants discern the primary value of their project in the findings of qualitative research—especially in the absence of solid expectations or hypotheses for the data—they typically engage in a more strict CQA process \cite{lazar_research_2017} or even grounded theory, as previously described. This approach fosters a deeper, more inductive, and nuanced understanding, consequently leading to insightful revelations. One participant explained the rationale behind this approach:}
\textit{"I think the reason [for following a strict CQA process in this project] maybe because first, we have more people and collaborators, and secondly, because this paper mostly depends on the qualities of the codes we've evolved. We wish it's a primary contribution, so it's like the difference in purpose [determines the methods and the strictness levels we use]." (P4, Ph.D. student with 4 years of QA experience).}

\added{In the context of a study that encompasses mixed contributions, participants tend to favor thematic analysis when they have clear expectations and a more structured framework in mind. This approach, encompassing steps from "Generate initial codes" to "Define themes" \cite{braun2006using, maguire2017doing}, places a greater emphasis on the testing and refinement of pre-defined themes than strictly following the step-by-step collaboration described above. One potential deviation observed is the utilization of a predefined codebook instead of initiating the coding process with open coding, which is a characteristic of the traditional CQA methodology. As explained by one participant:}
\textit{"For me, for example, if it's research mostly employing mixed methods, meaning you have both quantitative and qualitative data, in that case, if you have the quantitative research, then you use the qualitative to support this argument from another perspective. In that case, you can directly go for something like thematic analysis, and it's easy because you know what to expect." (P3, Ph.D. student with 4 years of QA experience).}

\subsubsection{Difficulties in Performing Collaborative Qualitative Analysis}

\paragraph{D1: Slow and time-consuming.} 
\label{sec:difficulty_slow}

The participants unanimously acknowledged that CQA is a time-intensive endeavor. The complete process of qualitative coding alone can span several weeks to several months for individual coders. When collaboration is involved, the duration is further extended due to "multiple rounds" of discussion and testing a comprehensive codebook \textit{(P4, Ph.D. student with 4 years of QA experience)}. As expressed by one participant: 
\textit{"It does take a lot of time. A 30-minute interview could be converted into seven or eight thousand words. I had to spend an hour or two to do independent coding, and another three hours to discuss." (P5, Ph.D. student with 2 years of QA experience).} 
Moreover, in cases where the coding process extends over a significant duration, coders often find it necessary to re-read the text during the reflection stage.
\textit{"The difficulty [of CQA] could be time-consuming. Especially if you do it twice, you have to read it again." (P1, a postdoc with 9 years of QA experience).}
Additionally, if one of the coders in the coding team works at a slower pace or struggles to keep up with the progress, it can result in an extension of the overall coding time \textit{(P3, Ph.D. student with 4 years of QA experience)}.

Yet, the time-consuming nature of the CQA process presents a valuable opportunity for junior researchers to actively engage and contribute, as they often have more flexibility and availability, allowing them to dedicate ample time to the rigorous coding process involved in CQA. For example, half of the participants (P1-P4) were involved in projects that fostered collaboration with junior researchers, including those with little to no prior coding experience. Remarkably, in these collaborations, the code suggestions from both groups were treated with equal importance and consideration. One participant explained their approach:
\textit{"Actually I work with some junior students. In this case, I think the first time we would encourage them to read the book chapters about how to conduct coding...After a few runs of demonstrations, in most cases, they become, you know, better QA [coders]." (P4, Ph.D. student with 4 years of QA experience)}

\paragraph{\added{D2: Coding bias and the struggle for independent coding}}

\added{In a strict CQA process, individual coders are expected to do coding independently from their own perspectives. However, the current QA software available, such as Atlas.ti, often poses challenges in merging codes when coders perform open coding independently in separate documents. This limitation can lead to coders relying on shared documents on Atlas.ti Web or Google Docs to label the data, benefitting from their easy and real-time data synchronization feature. }

\added{However, this shared feature has its drawbacks. The visibility of each other's codes may result in mutual influence, potentially biasing the coding process. This concern was reported by 6 out of 8 participants.
As one participant described:}
\textit{"So the main one [difficulty] is usually very hard to separately do the coding...if one person has coded, then the next person will see that person’s coding, which means that you are influenced by other coders’ coding. You cannot select codes you added or switch off all the coding coming from others' perspectives. The view gets very overlapped and then gets very confusing." (P2, postdoc with 7 years of QA experience)}.

Similarly, although P6 did not explicitly mention the issue of code bias, it was observed through her shared coding practices that she utilized Google Docs to collaborate with her colleagues on a shared page. This collaborative approach inadvertently allowed all collaborators to view and access the codes being generated.

\paragraph{D3: Trade-offs in utilizing QA software vs. traditional text editors} 
Most participants (6/8) favored traditional text editing tools like Google Docs and Excel over professional QA software like Atlas.ti or nVivo, finding the current features of text editors sufficient for their needs. \textit{"We use Excel because it just has all essential functions that we need, like filtering, data validation and removing duplicates." (P1, postdoc with 9 years of QA experience)}. 

This can also be seen as an outcome of the deliberation of weighing the learning costs against the potential benefits: QA software such as nVivo is described as having a "steep learning curve" \textit{(P4, Ph.D. student with 4 years of QA experience)}, yet it seemingly does not provide substantial advantages over conventional tools. Users anticipate more sophisticated features in return for their significant learning investment—like auto-grouping similar codes among coders. This deliberation often leads many to favor simpler tools that are easier to master and require a lower investment in learning. One Ph.D. student with 2 years of QA experience (P5) shared, \textit{"I also tried to use nVivo [to do collaboration]. I found that when using nVivo, I can not group together codes and text from three coders [automatically]. So later we just used Excel."} 

However, shifting to traditional text editors can pose challenges when handling large and diverse datasets, especially with an extensive codebook. As one participant noted, \textit{"The final codebook is very large, with a lot of themes, it is troublesome to read." (P5, Ph.D. student with 2 years QA experience).}
Moreover, users may forget their proposed codes in open coding, relying on "memory and perception" (\textit{P6, Master's student with 3.5 years of QA experience}) to generate the initial codebook. This often complicates subsequent stages of qualitative coding, requiring additional work, such as expanding the codebook or necessitating another round of CQA.

\paragraph{D4: Trade-offs of using a predefined codebook}
\label{sec:predefined}
Some participants (3/8) indicated that the practice of lead coders proposing a primary codebook for the team to follow could streamline the coding process. As one participant described, \textit{"I developed an initial codebook. I sent a copy to the two coders to let them do some coding on their own. Then after 1 week, they sent it back to me with their codes. Then we do a comparison to solve the disagreements." (P8, Master's student with 1.5 years of QA experience).} 

Yet, this could potentially restrict the benefits of CQA. P1, a postdoc with 9 years of QA experience, shared her opinion, \textit{"It sorts of restricts the categories. They only had to code the data in a certain way."}

However, some participants acknowledged situations where, despite recognizing it as less than "strict", they would adopt this strategy due to time constraints: \textit{"For my project, we have very limited time to do coding, then my collaborator and I do coding for the whole transcripts with the codebook I proposed. If I have enough time, I would definitely do a stricter qualitative analysis."(P4, Ph.D. student with 4 years of QA experience).}

\paragraph{D5: Interpretational variations and granularity challenges}
\label{sec:difficulty_level}
Interpretational divergence is a common occurrence among coders working on the same data, leading to distinct coding outcomes. P1, a postdoc with 9 years of QA experience, stated, 
\textit{"The difficulty comes after [coding] the first round codes. We always have different ways of interpreting things."}.
Furthermore, coders could assign codes with differing levels of specificity, causing extra work and necessitating further discussion to reconcile the codes: \textit{"My codes are broader, my partner is more specific. My codes usually cover the whole category. But hers give a bit more subcategories...the thing is I don’t want the subcategories. I just want all the codes to be the same categories."(P7, Master's student with 3 years of QA experience)}.

\subsubsection{Suggestions for AI-assisted CQA tools}
\paragraph{S1: AI generates code suggestions based on code history} The majority of our participants (6/8) express a desire for AI to provide code suggestions, with the most preferred source being their personal coding history.
\textit{"So if this content and that content suddenly matches up, it would be nice for AI to just assist me and say, `hey, you've done this before! Why don't you assign this code to this code that you have already assigned?' And I may not have remembered, because there were a lot of things I processed along the way." (P1, a postdoc with 9 years of QA experience).} 

P3, a Ph.D. student with 3 years of QA experience, expressed skepticism towards an AI system in which codes are derived from external sources, such as other projects or AI-generated suggestions: \textit{"Initially, the coding is a thinking process. I don't want to interrupt the [open coding] thing. I don’t want to be biased by somebody in the thinking process...Because any suggestion is biased. Once I finish initial coding, AI can suggest like, for example, then AI just shows like these are similar. These are different. Then I can use that feature similar to the initial coding. Then you refer my codes to feed the AI suggestions. I think that can improve the coding. If AI suggests [by its own], I don't trust anyway."}

\paragraph{S2: AI facilitates coding conflict detection}
\added{In our discussions with participants, we confirmed Zade et al.'s \cite{zade2018conceptualizing} previous analysis that a text selection analyzed by multiple coders could lead to 1) divergence: entirely distinct or even opposing interpretations, or 2) diversity: identical interpretations conveyed through different expressions, such as "not bad" and "well". Both scenarios necessitate coders to scrutinize the text, highlight the disparities and engage in a dialogue to reach a consensus on the final code. This process represents a significant time commitment.}

To handle the "diversity" conflict, our participants (4/8) anticipate that AI can play a crucial role in automatically detecting variations in codes assigned by different coders to the same text. This would prompt them to continuously reassess their coding choices in real-time throughout the coding process. As described by P7, 
\textit{"I think the collaboration is a matter of recommending what other collaborators have already done...for instance, one sentence, `oh the chatbot did not understand me.' Then my code is `chatbot is stupid', and the code of another coder is `chatbot has insufficient data'. AI may detect this difference [in real time]." (P7, Master's student with 3 years of QA experience).}

Moreover, P7 further explained that this conflict detection can take place during the coding process, fostering timely discussion among coders and eliminating the necessity to recall the meaning of their own codes or engage in manual one-by-one comparisons after coding. As a result, they would only need to allocate a lower amount of time to resolve discrepancies. This is particularly true when confronting with code "divergence":
\textit{"Then the difference may be detected. AI says, `So what's the difference? Why did you choose a different code? If you choose this code, can you give a reason why you choose this code?' Then I tell the other coder to come back, `guys. Let us discuss' (P7, Master's student with 3 years of QA experience).}

\subsection{Study Limitation}
\added{It is important to acknowledge that our findings may have limitations. One main limitation is the background of our participants, as all but one (P2) are coming from a technology-oriented background, as indicated in Table \ref{table:userstudy}. This could potentially impact our findings, as their projects often involve mixed methods and practices that may be influenced by their technological perspectives rather than perspectives and methods rooted in social science or psychology. 
Our study also includes participants with varying experience with CQA (experts and non-experts). Our goal here was to encompass a range of perspectives and experiences, allowing us to address both the learning issues associated with applying CQA and the inherent issues that arise in its implementation.}

\subsection{Discussion}
In practical applications, similar CQA steps are employed by our participants with slightly different forms, but the majority of practitioners primarily adhere to the six CQA steps proposed by Richards et al. \cite{richards2018practical}. 
Regarding the anticipated stages of AI integration, our findings support the conclusions drawn in the previous study conducted by Feuston et al. \cite{feuston2021putting}, that AI can be effectively incorporated into the inductive coding process. However, unlike previous research that primarily focuses on utilizing AI for pattern identification and data interpretation at this stage, our findings demonstrate an alternative use of AI: detecting coding conflicts between coders and facilitating collaborative interpretation and evolution of data among them.
Building upon the established CQA stages and the anticipated integration of AI, we have designed a study task for our evaluation of \name consisting of three primary phases: 1) open coding, 2) codebook formation through discussion, and 3) coding using the codebook.

Our findings also confirmed several noteworthy challenges and limitations associated with CQA. These include the time-consuming nature of the process~\cite{marathe2018semi, rietz2021cody}, difficulties in achieving consensus among coders~\cite{cornish2013collaborative, zade2018conceptualizing, drouhard2017aeonium, jiang2021supporting, feuston2021putting}, and the presence of software-related issues~\cite{hopper2021youtube, jiang2021supporting, feuston2021putting}.
In particular, we have identified a new limitation related to coding bias arising from the challenges of independent coding. This limitation is particularly observed in the current CQA software that facilitates real-time collaboration and coding on a shared document among coders. For instance, in platforms like Google Docs and Atlas.ti Web, users have visibility into each other's codes while co-coding. This visibility negatively impacts their willingness to use CQA software as it introduces a potential bias in the coding process. 

\added{In an effort to mitigate this limitation, we propose a solution that involves leveraging AI as an intermediary between the two coders. Instead of directly revealing each other's codes, our system retrieves and analyzes the coded data from both coders in real time. Incorporating the distinct perspectives of each coder, our AI system generates code suggestions that serve as indirect indicators of coding differences, which facilitate awareness of differences among the coders. This awareness encourages them to reconsider and reflect upon their own codes during the coding process, rather than solely during post-coding discussions. By doing this, we aim to minimize bias in the existing coding process when using traditional real-time collaboration software such as Google Docs, Atlas.ti Web, and others.}

\section{System Design}
Based on the insights from the interviews, we have proceeded to design and implement a platform that harnesses the power of AI to facilitate the CQA process.

\subsection{Design Consideration}
\label{sec:req}

\begin{enumerate}
    \item \added{Independence and convenience: To keep both independence and convenience (D2), each coder is assigned a separate web page with the same data, but allows access to others' coding results by simply clicking a link rather than navigating through the cumbersome process of exporting and importing that current non-web software necessitates \cite{marathe2018semi}.} 

    \item \added{Lower learning curve: To mask the complexity of AI and flatten the learning curve (D3), our system is designed to emulate familiar platforms like Google Docs, specifically its "comment" function. This design enables the addition and removal of codes in a way that users can readily comprehend and navigate. Moreover, the web page or coding interface allows data selection at various granularity levels for code addition, similar to existing commercial QA software.}
    
    \item \added{List of code suggestions: Our approach to collecting coding history and generating code suggestions for coders has the potential to yield significant time savings and reduce manual effort (D1). In accordance with Rietz et al.~\cite{rietz2021cody}, our approach includes providing multiple AI suggestions, each accompanied by a confidence level. This aims to bring attention to the inherent uncertainty of codes and curtails the potential risk of thoughtlessly adopting these suggestions \cite{chen2016challenges, jiang2021supporting, chen2018using}.}

    \item \added{User autonomy: Code suggestions are revealed only upon user request, emulating the natural coding process. This fosters active thinking before viewing suggestions, minimizing the risk of being unintentionally guided in unwanted directions \cite{jiang2021supporting} by direct exposure to others' codes (D2). This is beneficial as it encourages independent thinking, a tool to combat groupthink \cite{janis2008groupthink}.}

    \item \added{Collaboration: The coding histories of both coders are comparatively analyzed and subsequently input into an AI classification model. This model then generates a range of AI suggestions for coders upon request (S1). It not only provides a reference point to users based on their own coding history but also integrates their partner's history. Through this design, our goal is to foster awareness of disparities among coders, thus enhancing their understanding and reflection of the data. This becomes particularly beneficial when addressing coding styles or logic conflicts \cite{akpinar2009role} within a group.}

    \item \added{Usage in inductive coding: Beyond the system, instead of enforcing a rigid, predefined codebook that could potentially constrain coders (D4), we place importance on the dynamic nature of the inductive coding process in the evaluation.}
\end{enumerate}

The specific design components of our prototype, \name, are elaborated in the following subsections.

\subsection{Interface} 
\begin{figure}[!t]
  \centering
  \includegraphics[width=\linewidth]{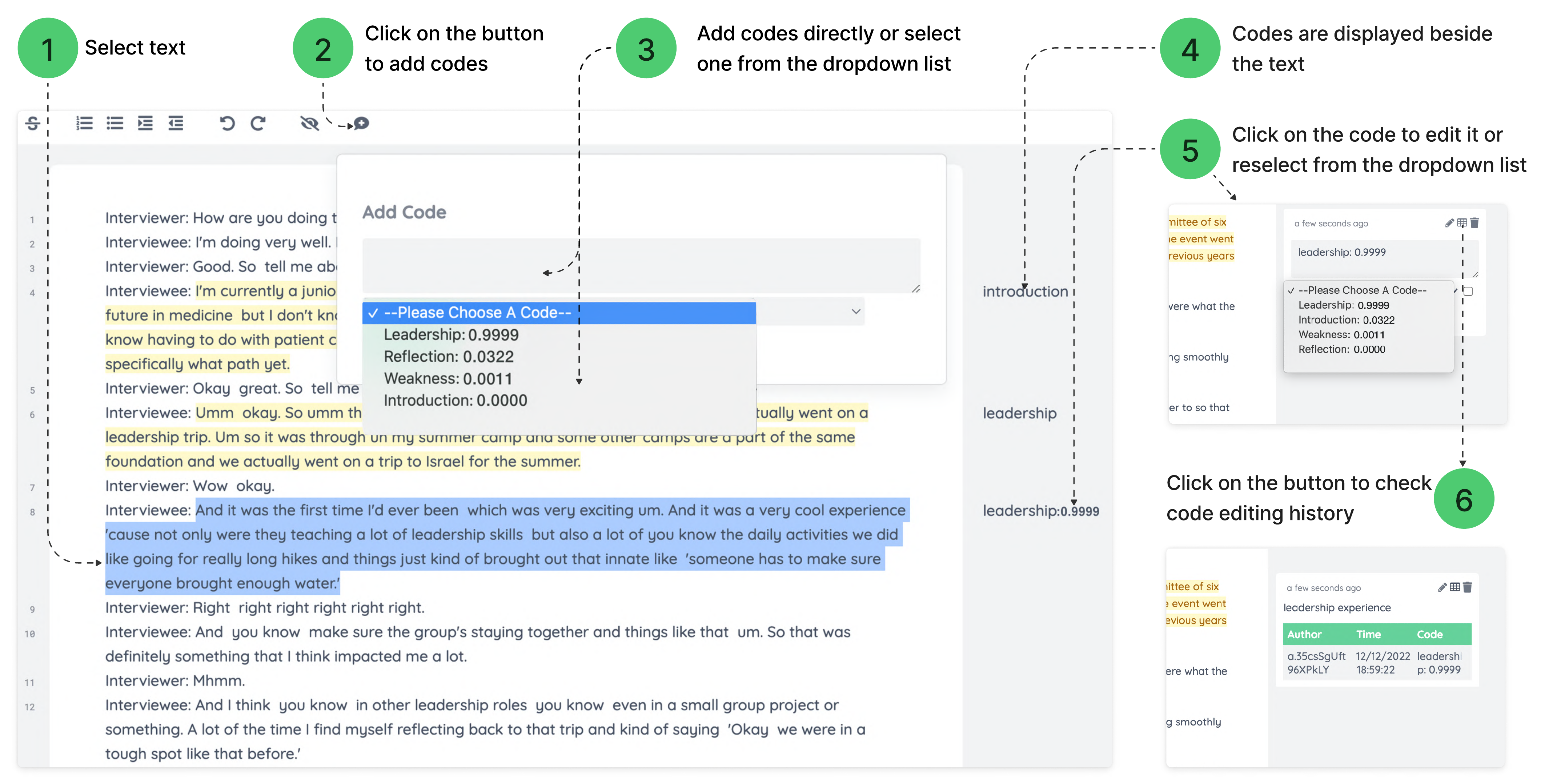}
  \caption{The \name interface. Creating a code: (1) Users select the text of significance and (2) click the comment button to add codes. (3) Users can add codes directly or select one from the dropdown list. (4) The created code is shown beside the selected text. (5) Click on the code to edit it directly or reselect codes. (6) Click on the button to check the code editing history.}
  \label{fig:interface}
\end{figure}

The interface (see Figure \ref{fig:interface}) is built on two components: 1) \textit{Etherpad}\footnote{\url{https://etherpad.org/}}, an open-source collaborative text editor that supports multiple users editing text in real-time \cite{amiryousefi2021impact, bebermeier2019use, goldman2011real}, and 2) its plugin, \textit{ep\_comment\_pages}\footnote{\url{https://github.com/ether/ep_comments_page}}, which allows for adding comments beside the text.
To create a code, users select the text of significance $\rightarrow$ click on the "comment" button $\rightarrow$ type in the code OR select from a list of AI-generated code suggestions $\rightarrow$  press "save". 
The interface also provides features for coders to review the code history, re-edit, and delete the code in case they change their minds.
Additionally, the customized version of \textit{ep\_comment\_pages} offers code suggestion lists containing a maximum of five codes when requested by the user. These codes are ranked by their confidence level, which ranges from 0 to 1.

\begin{figure}[!t]
  \centering
  \includegraphics[width=\linewidth]{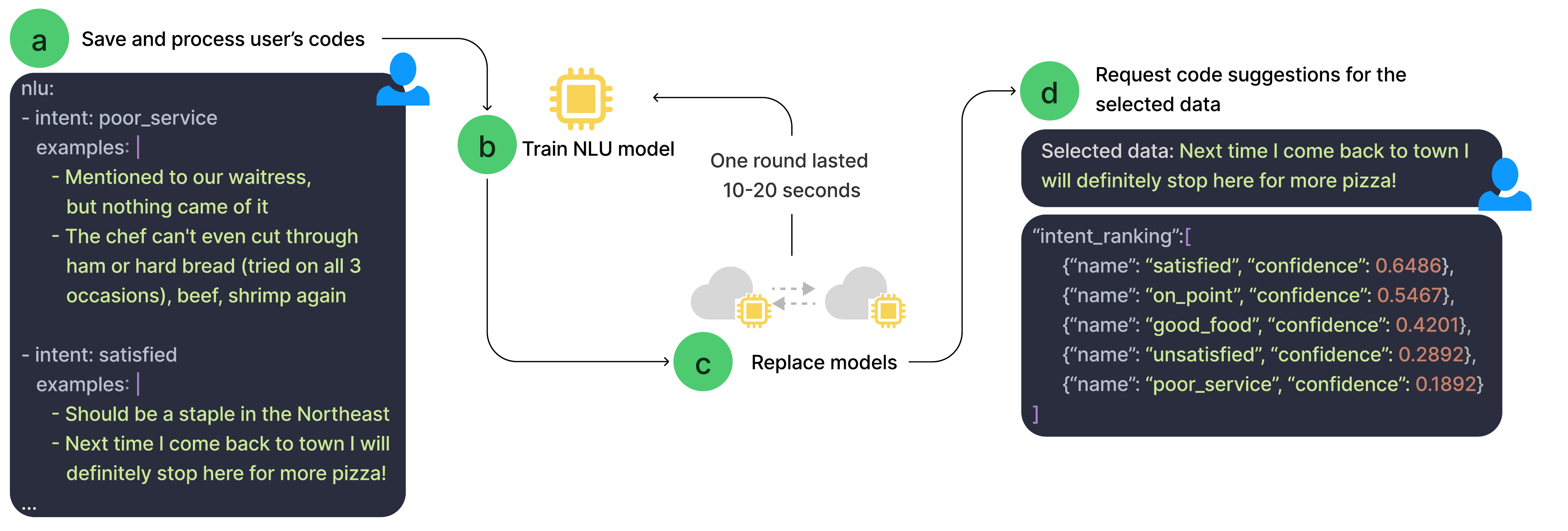}
  \caption{The pipeline for training and updating the \name model, designed to facilitate code suggestion requests. (a) Save and process user's coding data: This step involves saving, retrieving, and processing each coder's coding data. The retrieved data is then used to generate an \textit{nlu.yml} file, which contains coded text (i.e., examples) and codes (i.e., intent). 
  (b) Train NLU Model: This process trains a new model using the updated \textit{nlu.yml}, which takes about six seconds or longer, depending on the coding data size.
  (c) Replace models: This process substitutes the old model with the newly trained one, approximately requiring four seconds per model.
  (d) Request code suggestions: The user requests code suggestions from the server. Initially, \name requests code suggestions from server1. If this fails, the request is then rerouted to server2, thereby sustaining the impression of continuously updated code suggestions for the user.}
  \label{fig:training}
\end{figure}

\subsection{AI Model}
\label{sec:AImodel}
The AI model harnesses the Rasa NLU  framework\footnote{\url{https://rasa.com/docs/rasa/}}, to generate code suggestions upon a user request. Previously, Rasa \cite{bocklisch2017rasa, cao2023understanding} has been deployed in the domain of HCI for handling conversations in several prototype models \cite{porfirio2019bodystorming}.
Specifically, we employ a recommended NLU pipeline from Rasa to train an NLU classification model\footnote{\url{https://rasa.com/docs/rasa/tuning-your-model/\#configuring-tensorflow}}. This pipeline incorporates multiple components: \texttt{SpacyNLP}, \texttt{SpacyTokenizer}, \texttt{SpacyFeaturizer}, \texttt{RegexFeaturizer}, \texttt{LexicalSyntacticFeaturizer}, two instances of \texttt{CountVectorsFeaturizer}, and \texttt{DIETClassifier}.

Within this pipeline, the SpacyNLP language model `en\_core\_web\_sm'\footnote{https://spacy.io/usage/models} is selected for training efficiency consideration, comparing to larger pre-trained language models utilized in the SpacyFeaturizer\footnote{It should be noted that our coding materials in the evaluation consist solely of simulated job interview transcripts and do not encompass any specialized domain knowledge.}.
Moreover, the DIET (Dual Intent and Entity Transformer) Classifier \cite{bunk2020diet} is selected for its capability to perform multi-class classification. The NLU pipeline operates on a computer equipped with Ubuntu 20.04, Tensorflow (2.6.1), CUDA (11.2), and an Nvidia GPU Quadro K2200 graphics card, in conjunction with installed software like Rasa (3.0), Node.js (17.2.0), and MongoDB (5.0.4).

The AI model, trained on users' coding histories, may not provide suggestions for the initial few requests due to a lack of historical data. However, as the data pool expands, it gains the capability to generate up to five code recommendations for each request, sorted according to their respective confidence levels. The DIET  Classifier computes these confidence levels, indicating the cosine similarity score between predicted labels and text\footnote{\url{https://rasa.com/docs/rasa/components/\#dietclassifier}}.

\subsection{Training and Updating Pipeline}
The training and updating pipeline is shown in Figure \ref{fig:training}.

\subsubsection{Save and Retrieve Data}
\label{sec:save_data}
Each user's coding data is individually stored in the database. To identify conflicts, the codes of each coder are compared with both their own and their peers' coding histories. The codes are subsequently grouped and deduplicated, readying them as inputs for the NLU pipeline.
For example, if two sentences are labeled with the same code, they are grouped into one "intent" (akin to the "class" concept in Machine Learning, and equivalent to "code" in this work) in the Rasa NLU data file, \textit{nlu.yml}. If a single sentence is coded with two distinct codes by two coders, it serves as an "example" for both "intent" in the Rasa NLU file. If two codes convey similar meanings but have different expressions, they are interpreted as two separate "intent" (in the current version of \name). It should be noted that this process may slightly vary under the four conditions outlined in Section \ref{sec:study_design}.

\subsubsection{Train and Reload NLU Models in Real Time}
Firstly, the \textit{nlu.yml} file is fed into the NLU pipeline for automatic training of the new NLU model. Secondly, the trained model is promptly uploaded to the Rasa server via HTTP API, replacing the previous model.
Lastly, we configure two Rasa Open Source servers to run on ports 5005 and 5000, respectively. 
These servers act as buffers for user requests, utilizing server swapping.
Users have the capability to request code suggestions through HTTP from either of the two servers. In case \name fails to receive a response from one server due to the server's ongoing model update process, it swiftly switches to the other server as an alternative.
The complete pipeline typically requires approximately 10 to 20 seconds or longer, depending on the size of the coding data. Users experience a seamless process without any interruptions caused by model updates.
\section{User Evaluation Design}
\label{sec:study_design}
We proposed four collaboration methods for using \name (see Figure \ref{fig:collaboration}). To ensure a fair evaluation, we focus on novice users who primarily participate in the CQA process (see section \ref{sec:difficulty_slow}) and undergo requisite coding training. We further strive for an evaluation of \name across varied collaboration modes in a between-subject study, considering they may originate from a similar starting point.
This user study has been approved by our university's IRB.

\subsection{Task}
We simulated a CQA task encompassing three main steps: independent and open coding, codebook formation through discussion, and codebook application. Due to the multiple factors involved in our study, we simplified the study by focusing on the CQA process involving two users. To determine the optimal duration for each phase and the overall study, we conducted a pilot test with eight participants (5 females, 3 males, mean age = 24.7).

\begin{figure}[!t]
    \centering
    \includegraphics[width=\linewidth]{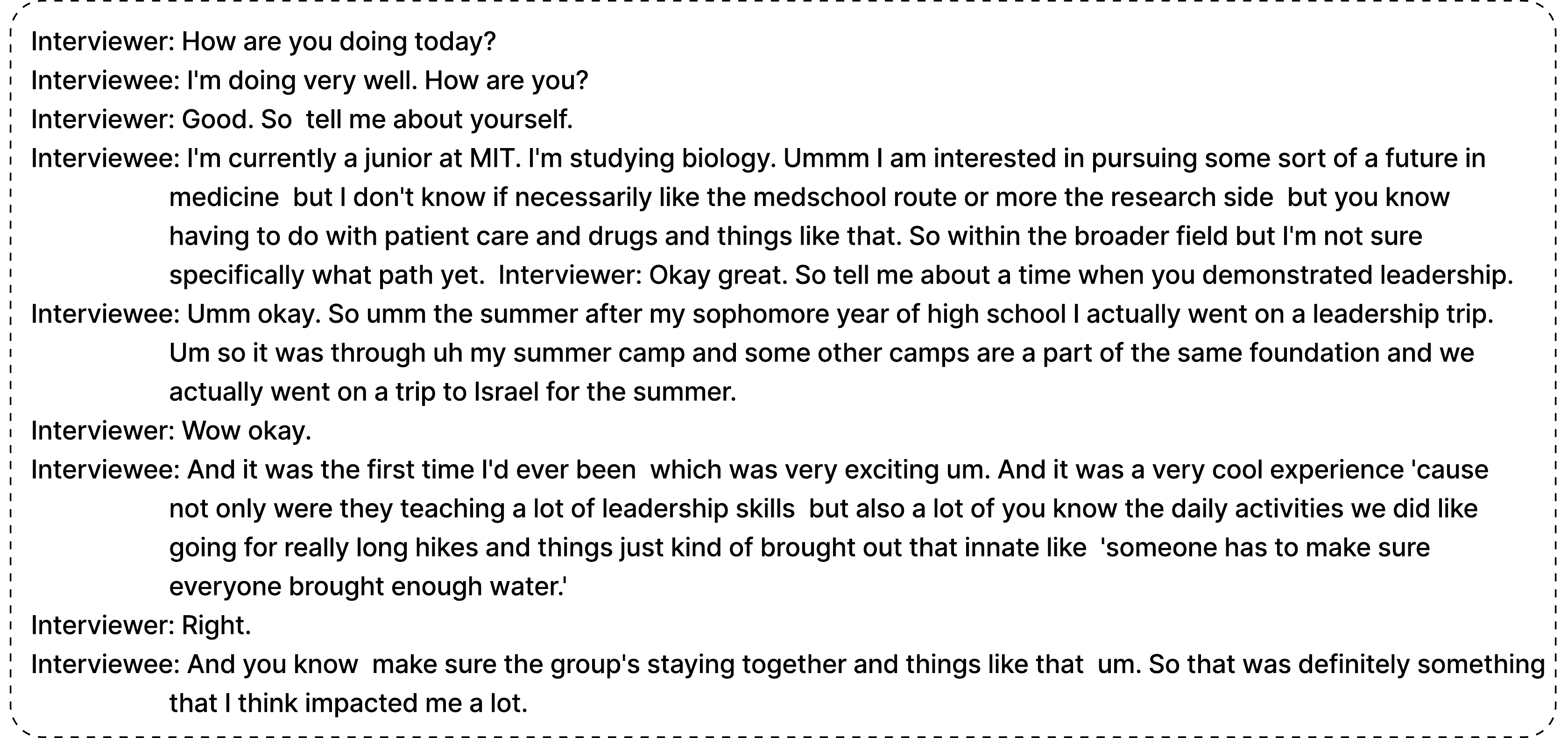}
    \caption{One sample interview transcript in coding task.}
    \label{fig:samplema}
\end{figure}

We selected mock interview transcripts from an open-source dataset~\cite{naim2015automated} as the text materials to be coded in our study. This dataset covers various general topics, including leadership, personal weaknesses, and challenging experiences, which are familiar and accessible to most users.
To ensure control over potential effects on both the AI model and user understanding, we specifically chose three transcripts that exhibit better clarity and coherence, each consisting of approximately 1000 words. Part of a sample interview transcript is provided in Figure \ref{fig:samplema}.

\subsection{Independent Variables (IVs) and Conditions}
To understand the impact of AI on human-to-human collaboration in CQA, we identified three factors for our study:

\begin{enumerate}
    \item AI \{With, Without\}: Whether or not AI (the NLU model) is applied to provide code suggestions. This IV aims to understand whether the use of AI affects CQA performance.
    \item Synchrony \{Synchronous, Asynchronous\}: Whether or not two coders do coding in real-time and simultaneously. This IV aims to understand whether CQA performance is achieved similarly in synchronous and asynchronous modes.
    \item Shared Model \{With, Without\}: Whether or not two coders use a shared NLU model to request AI suggestions. The shared model can collect coding history from both coders and be trained on it to provide code suggestions. Without a Shared model, each coder can only get AI suggestions based on his/her own coding history. This IV aims to understand if a shared model affects the CQA performance.
\end{enumerate}

\begin{figure}[!t]
\centering
\includegraphics[width=0.85\linewidth]{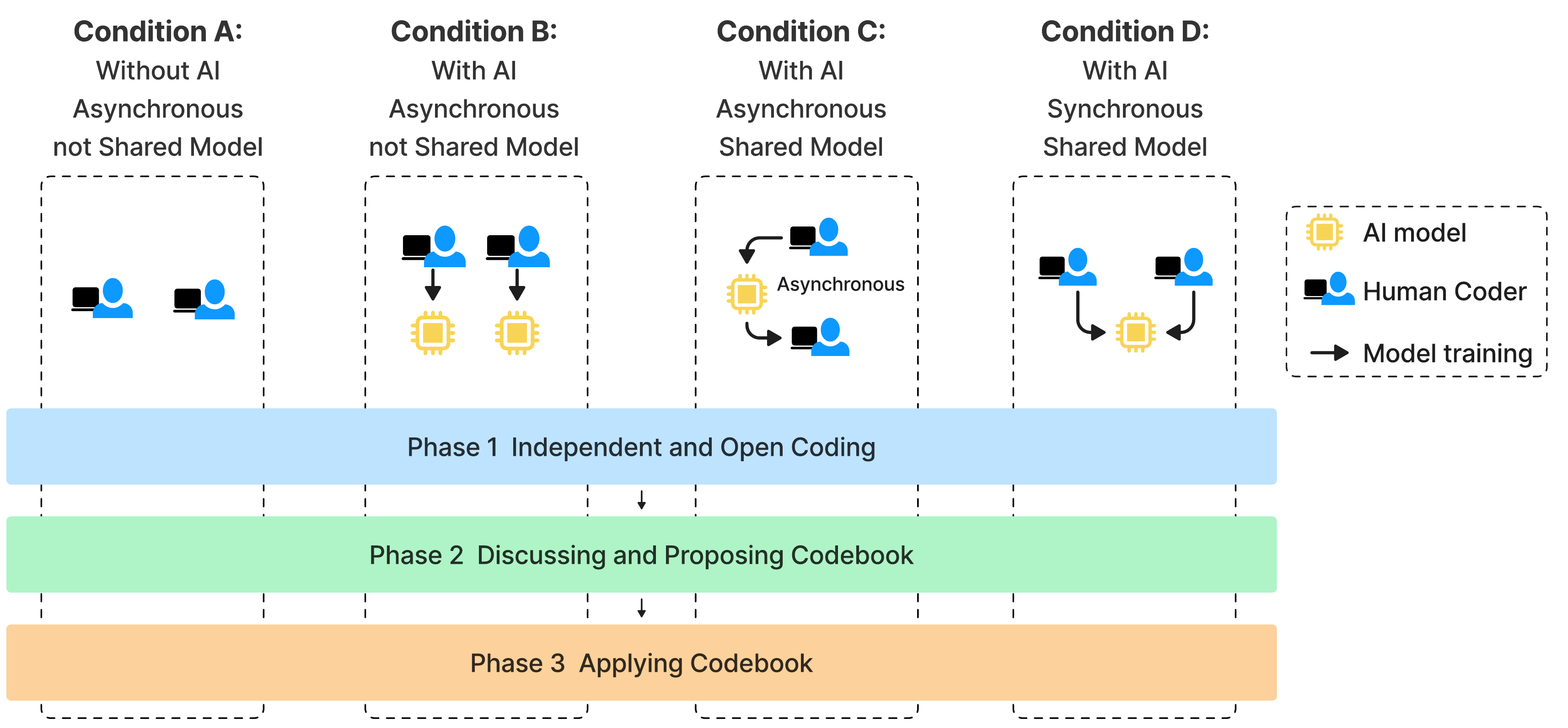}
\caption{Four Approaches to Collaboration in Qualitative Analysis.
\condA (Traditional Coding): Both coders independently apply codes.
\condB: Each coder applies codes using their respective NLU models.
\condC: The coders apply codes asynchronously with a shared NLP model. Coder1 begins the process, during which the NLP model trains and offers real-time AI suggestions. Once Coder1 completes the task, Coder2 commences with coding.
\condD: The coders apply codes synchronously with a shared NLP model.}
\label{fig:collaboration}
\end{figure}

We combined these factors and removed meaningless and duplicate combinations, resulting in four final conditions (A-D) for the collaboration (see Table~\ref{table:factors}).
The final conditions are shown in Figure~\ref{fig:collaboration}.

\begin{table}[htbp]
\caption{The combinations of the three factors, which are not entirely independent of each other. For instance, the absence of AI renders the Shared Model factor irrelevant, making conditions C7 and C8 nonsensical. Synchronous coding is only applicable in the presence of a Shared Model, as the order in which coding occurs is inconsequential without a shared model. Thus, certain conditions become identical due to the absence of a shared model, namely C1 and C2, as well as C3 and C4.}
\label{table:factors}
\scalebox{0.8}{\begin{tabular}{@{}lllllllll@{}}
\toprule
Combination  & C1 & C2 & C3 & C4 & C5 & C6 & C7 & C8 \\ \midrule
AI           & $\times$  & $\times$  & \checkmark  & \checkmark  & \checkmark  & \checkmark  & $\times$  & $\times$  \\
Synchrony    & $\times$  & \checkmark  & \checkmark  & $\times$  & $\times$  & \checkmark  & $\times$  & \checkmark  \\
Shared Model & $\times$  & $\times$  & $\times$  & $\times$  & \checkmark  & \checkmark  & \checkmark  & \checkmark  \\ \midrule
Condition    & A  & A  & B  & B  & C  & D  & $\times$  & $\times$  \\ \bottomrule
\end{tabular}}
\end{table}

\condA (Traditional Coding): In Phase 1, each coder independently applies codes to the first two interview transcripts using distinct web pages. In Phase 2, the coders convene to establish a shared codebook. During Phase 3, they independently apply codes to the third transcript based on the codebook. As new codes are entered, the model undergoes automatic updates.

\condB: In Phase 1, each coder independently applies codes to the first two interview transcripts, each utilizing an individual model on separate web pages. These two independent models are automatically trained during the coding process, with the training data sourced from their individual code histories. In Phase 2, the coders convene to formulate a shared codebook. During Phase 3, they independently apply codes to the third transcript based on the codebook, with the model undergoing automatic updates as new codes are entered.

\condC: In Phase 1, coder1 independently assigns codes to the first two interview transcripts. The model is concurrently trained and offers real-time code suggestions as the coding process unfolds. Once coder1 completes the task, coder2 commences coding on a separate web page. Initially, the code suggestions are entered by coder1's coding history. Over time, as coder2's codes are introduced, they are incorporated into the suggested codes. In Phase 2, the coders collaborate to create a shared codebook. During Phase 3, they independently assign codes to the third transcript using the codebook, triggering automatic updates of the model as new codes are inputted.

\condD: In Phase 1, both coders independently assign codes to the first two interview transcripts, each working on a separate web page. As the coding process progresses, \name accesses their individual code histories and automatically trains the model 3-6 times per minute. Upon request, it then provides code suggestions for both coders. In Phase 2, the coders collaborate to develop a shared codebook. During Phase 3, they independently apply codes to the third transcript using the codebook. The model is automatically updated when new codes are inputted.

\subsection{Participants}
A total of 64 participants (41 females, 23 males), ranging from 18 to 57 years old (mean=25.3, median=23), were recruited through our university's email system and public channels on Telegram targeting other universities within Singapore. All participants were native English speakers and reported no prior experience in qualitative analysis. In accordance with our university and national guidelines for participant reimbursement, each participant received an hourly compensation of 10 Singapore Dollars.

\subsection{Procedure}
\label{sec:study_design:procedure}
\begin{figure}[!t]
    \centering
    \includegraphics[width=0.8\linewidth]{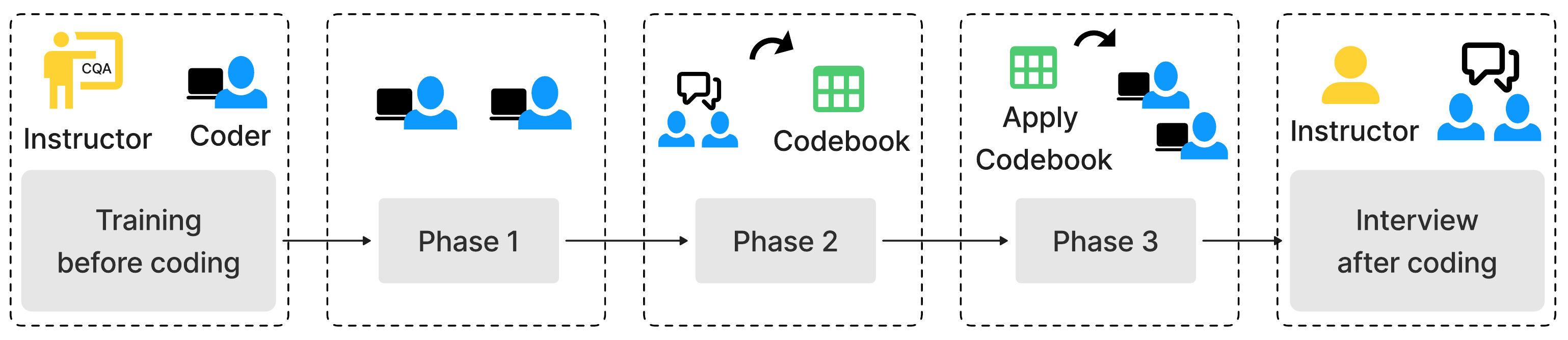}
    \caption{Study procedure. Both coders underwent training in CQA, prior to the formal coding process. 
    \textbf{Phase 1 (Independent and Open Coding)}: In this phase, both coders individually performed coding for two interview materials, following the assigned study setup ($\leq$20 minutes). 
    \textbf{Phase 2 (Discussion and Codebook Formation)}: During this phase, the two coders engaged in discussions to collaboratively create a structured codebook using Google Sheets ($\leq$40 minutes).
    \textbf{Phase 3 (Application of the Codebook)}: In this phase, the coders independently applied the codes from the agreed codebook during their individual coding sessions ($\leq$10 minutes). At the end of each phase, participants were required to complete a survey and interview ($\approx$5 minutes).}
    \label{fig:process}
\end{figure}

The final study process was structured based on the setup depicted in Figure \ref{fig:process}. A total of 64 participants were divided into 32 pairs, with each study group comprising 8 pairs. The allocation of participants to the four study groups was random, corresponding to the four conditions specified in Figure \ref{fig:collaboration}.

To ensure participants had a thorough understanding of the task, we conducted an approximately 15-minute training session prior to the study. The training session consisted of two key components:

\begin{enumerate}
    \item Instruction: Participants received instruction from an instructor's explanation that demonstrated how to use \name for selecting the text, adding codes in Phase 1, creating a codebook in Phase 2, and applying it in Phase 3. 
    \item Training Tutorial and Q\&A: Participants were shown a video explaining the principles of qualitative coding, and they also received a written tutorial explaining the concepts of qualitative coding, codes, subcodes, and accompanying examples. A question and answer session was held to address any queries and ensure clarity.
\end{enumerate}

\added{Following the training session, participants engaged in the three phases as illustrated in Figure \ref{fig:process}. After the study, we also gathered demographics information and invited participants to provide feedback through an interview regarding their experience with the system. The interview encompassed questions about their experience using the software, focusing on individual and collaborative coding challenges, as well as their attitude towards \namef features. 
A detailed list of questions can be found in the study protocol within the Appendix \ref{appendix:study_protocol}. With the participants' consent, we made audio recordings of the post-interviews to facilitate subsequent analysis.
This analysis allowed us to gather valuable insights and perspectives on the user experience of the tasks. }

In addition, we implemented the following measures to ensure the tasks were performed effectively:

\begin{enumerate}
    \item Time reminders: Regular reminders were provided to participants to keep them aware of the remaining time. For example, notifications were given when there were 15 minutes left and 5 minutes left in the allotted time frame.
    \item Coding quality check: We monitored their codes and selected text to address any questions, issues, or potential misunderstandings that participants may have encountered during the task. However, we made a conscious effort to minimally interfere with their coding process to maintain the integrity of their work. 
    \item Encouragement for diverse codes: Actively encouraged participants to generate a wide range of diverse codes during the coding process. 
\end{enumerate}

\subsection{Dependent Variables (DVs)}
\subsubsection{Coding Time} 
This DV quantifies the time taken for each of the three phases of the study (see Figure \ref{fig:process}). An approximate 90-minute time limit was enforced to regulate the study duration for each pair. Specifically, the time allotment for each phase is as follows: Training: $\approx$ 15 minutes; Phase 1: $\leq$ 20 minutes; Phase 2: $\leq$ 40 minutes; Phase 3: $\leq$ 10 minutes; Post survey and interview: $\approx$ 5 minutes.
This structure was established in light of our pilot studies, during which we observed that most participants were able to satisfactorily complete the coding task. However, the actual coding time usage may deviate from the estimate. We assessed the actual time used in each phase for further analysis.

\subsubsection{Inter-rater reliability (IRR)} 

IRR is a metric that gauges the level of consensus among multiple coders \cite{kurasaki2000intercoder}. In our study, we specifically employ Cohen's kappa ($\kappa$), a measure designed to compute agreement between two coders \cite{mchugh2012interrater}. The computation of $\kappa$ encompasses both Phase 1 and Phase 3.

\subsubsection{Code Diversity}
This DV quantifies the diversity of proposed codes, a factor that is intimately linked to coding quality. To measure this, we count the number of \textit{unique} codes—also referred to as first-level codes—and subcodes, or second-level codes. Here, "unique" signifies that variations of a similar meaning are counted as a single code.

\subsubsection{Code Coverage} 
This DV quantifies the degree of overlap between the individual coders' codebooks and the merged codebook, at both the first and second coding levels. This applies to the initial codebook from Phase 1 and the users' proposed codebook in Phase 2.
The merged codebook is iteratively developed by consolidating common codes from the 32 codebooks created during Phase 2.

The formula used for the calculation is:

$Code\ Coverage = \frac{\text{|$Coders'\ Codebook \cap Merged\ Codebook$|}}{\text{|$Merged\ Codebook$|}}$,

where $\cap$ represents the intersection of the two codebooks, and "| |" signifies the number of codes.

\subsection{Data Analysis}
\label{sec:study_design:data_analysis}

\subsubsection{Step 1: Data Integrity and Quality Checking}
Following data collection, our next step was to verify the integrity and quality of the collected data:

\begin{enumerate}
    \item 85\% Completion Rate: Participants should complete coding for more than 85\% of the provided data within the given time\footnote{Our time regulations were established based on pilot studies, during which most native speaker participants were able to complete the coding tasks. This is designed to prevent the study from becoming overly lengthy, which could lead to fatigue and a subsequent loss of focus and motivation.};
    \item Task Correctness and Active Collaboration: We examined whether participants performed tasks correctly and collaborated actively. Instances of extremely low code diversity, such as using a single broad code like "Experience" to describe all examples in the interview transcripts, were a cause for concern. This lack of diversity suggested an inability to form a quality codebook. Additionally, pairs that were not willing or able to engage in productive discussion, choosing instead to develop their own individual codebooks, were noted.
\end{enumerate}

Four pairs that did not meet these two criteria were omitted and replaced with new participants to maintain the data integrity and quality.

\subsubsection{Step 2: Generating Initial Codebooks}
\label{sec:initial_codebook}
\added{Phase 1 and Phase 2 represent the "pre-discussion" and "post-discussion" stages, respectively. In the pre-discussion stage, the codes present a higher degree of variation. Following the discussion stage, these varied codes have been deliberated and consolidated, with differing variants merged. }

\added{In order for us to assess the IRR, code diversity and code coverage, two authors then manually formed initial codebooks for each pair. These codebooks were specifically designated for assessing the aforementioned DVs, and were neither shared with nor used by the participants at any stage during the experiment. They first coordinated to merge codes of similar meaning, adhering to the following criteria and steps: codes conveying similar meanings but expressed differently were treated as second-level codes; subsequently, the authors collaborated to propose a corresponding first-level code for each central meaning in the initial codebook.}
For example, "\texttt{Introduction of Leadership Experience}", "\texttt{Description of Leadership Experience}", and "\texttt{Application of Leadership}" serve as three second-level codes that fall under one first-level code: \texttt{Leadership}. For more details, please refer to Table \ref{tab:initialcodebook}.

\begin{table}[htbp]
\caption{Part of a sample of the initial codebook (Phase 1). 
Each row containing second-level codes is counted as a single first-level code. This codebook demonstrates a "Code Diversity" of 5 first-level codes alongside 10 second-level codes.}
\label{tab:initialcodebook}
\scalebox{0.7}{\begin{tabular}{|l|lll|}
\hline
\textbf{\begin{tabular}[c]{@{}c@{}}First-level\\ Code\end{tabular}} & \multicolumn{3}{c|}{\textbf{Second-level Code}} \\ \hline
Career Goal & \multicolumn{1}{l|}{Personal introduction and future (career) goals} & \multicolumn{1}{l|}{Choosing of (academic and career) route} &  \begin{tabular}[c]{@{}l@{}} Not very sure about\\future (career) goal \end{tabular}\\ \hline
Personal Interest & \multicolumn{1}{l|}{Personal introduction and interest area.} & \multicolumn{1}{l|}{Interest in oil fossil fuels} &  \\ \hline
Leadership & \multicolumn{1}{l|}{Introduction of leadership experience} & \multicolumn{1}{l|}{Description of leadership experience.} & \begin{tabular}[c]{@{}l@{}}Application of leadership \end{tabular} \\ \hline
Teamwork & \multicolumn{1}{l|}{Intro of working with team on big project} & \multicolumn{1}{l|}{} &  \\ \hline
Initiative & \multicolumn{1}{l|}{Shows initiative} & \multicolumn{1}{l|}{} &  \\ \hline
\end{tabular}}
\end{table}

\begin{table}[htbp]
\caption{Part of a sample codebook, which is formulated from the 27-minute discussion in Phase 2 between participants P27 and P28 under \condD. The "First-level Code" column represents the first-level codes generated during this discussion. The "Second-level Code" column, on the other hand, contains codes proposed by them in Phase 1. The "Example" column showcases selected segments of the original text.}
\label{table:sampleco}
\scalebox{0.85}{\begin{tabular}{|l|l|l|}
\hline
\textbf{First-level Code} & \textbf{Second-level Code} & \textbf{Example} \\ \hline
\multirow{3}{*}{Interest and goals} & \multirow{2}{*}{Uncertain about the future} & \begin{tabular}[c]{@{}l@{}}Ummm I am interested in pursuing some sort of a future in\\ medicine but I don’t know if necessarily like the med school\\ route or more the research side but you know having to do\\ with patient care and drugs and things like that. So within\\ the broader field but I’m not sure specifically what path yet\end{tabular} \\ \cline{3-3} 
 &  & \begin{tabular}[c]{@{}l@{}}I think for me where I’m where I am at this point where I’m\\ deciding between sort of going the medical school route or\\ the research route;\end{tabular} \\ \cline{2-3} 
 & \begin{tabular}[c]{@{}l@{}}Show interest in\\ alternative energy\end{tabular} & \begin{tabular}[c]{@{}l@{}}I'm very interested in energy applications so um from alternative\\ energy to more traditional sources so basically oil and fossil fuels.\\ Um and kind of optimizing that industry I think there's a lot of\\ potentials there so that’s where my main interest is.\end{tabular} \\ \hline
\end{tabular}}
\end{table}

\subsubsection{Step 3: Measuring DVs}
\label{sec:dvs}
We evaluated various DVs, including \ctime, \irr, \diver, and \cover, throughout the three phases.

In terms of \ctime, we concluded each phase as soon as coders exceeded the time limit. All but four pairs completed the coding task within this limit – two in \condB, one in \condC, and one in \condD. However, we still consider these four groups as "completed" due to their achievement of a minimum 85\% completion rate.

For \irr, we segmented the complete interview data into sentences $\rightarrow$ represented codes numerically as "0" (for sentences without codes), "1", "2", "3", etc.
$\rightarrow$ Cohen's Kappa ($\kappa$) was calculated for first-level codes in Phases 1 and 3, as Phase 2 was a discussion session without new coding data. Due to the considerable variability, second-level codes were excluded from IRR computation as it was infeasible.

For \diver and \cover, both first-level and second-level codes from the initial codebook and the proposed codebook were incorporated into the computing process.

Moreover, we conducted a thematic analysis \cite{braun2006using, maguire2017doing} for the interview audio transcripts, given that most data align with the structure provided by the interview questions.

\subsubsection{Step 4: Statistical Analysis}
We conducted a non-parametric analysis on the quantitative data (\ctime, \irr), given concerns about the normality of the distribution of the data collected. Consequently, Kruskal-Wallis test was employed to identify the main effect, and Mann-Whitney U-test was used for pairwise comparisons.

\section{Quantitative Results}

\subsection{Coding Time}

\subsubsection{Total Time}
\begin{figure}[!htb]
    \centering
    \includegraphics[scale=0.65]{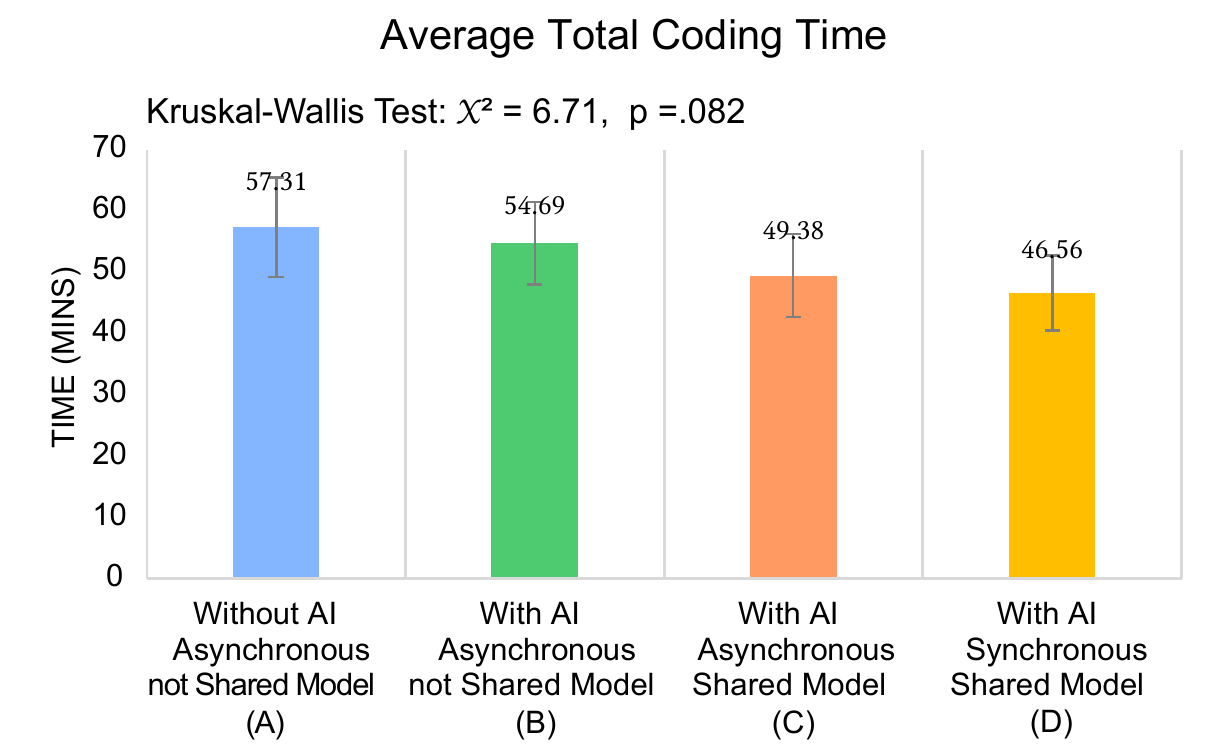}
    \caption{Average Total Coding Time for Each Condition (A, B, C, and D). Error bars show .95 confidence intervals. A Kruskal-Wallis test showed no main effect.}
    \label{fig:totalcodingtime}
\end{figure}

\begin{figure}[!htb]
 \centering
 \includegraphics[width=\textwidth]{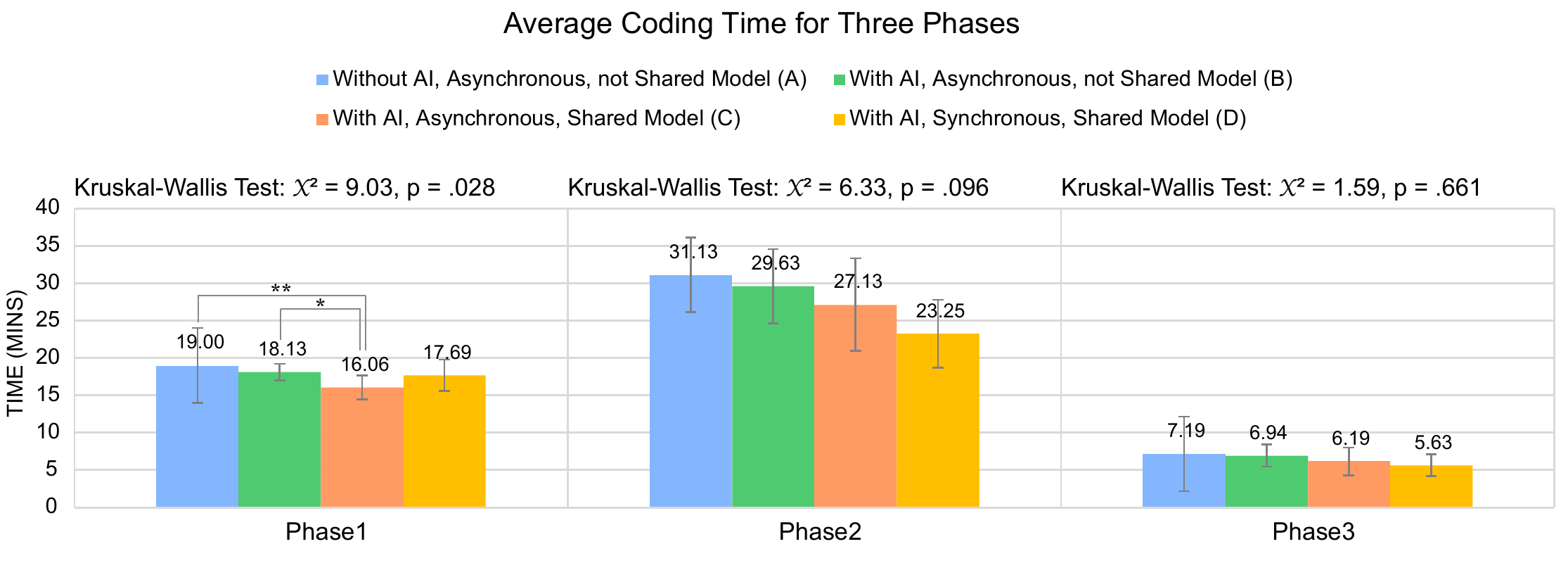}
 \caption{
 Average Coding Time for Three Phases. Error bars show .95 confidence intervals. We report the results of the individual Kruskal-Wallis tests, and, if necessary,  pairwise comparisons, where $*: p < .05, **: p < .01$.}
 \label{fig:separatecodingtime}
\end{figure}

The average total time for each study group and the average used time for three phases is shown in Figure~\ref{fig:totalcodingtime} and \ref{fig:separatecodingtime}. A Kruskal-Wallis test \textbf{did not reveal any main effect} of the conditions on the total time to complete the study (${\chi}^2_{(3)} = 6.71, p=.082$). In general, \condD was the fastest ($M=46.56\ mins$), followed by \condC ($M=49.38\ mins$), \condB ($M=54.69\ mins$) and \condA ($M=57.31\ mins$).
The time for individual phases is however more informative to understand the potential effect of AI on CQA performance, therefore, we also analyzed the time for each phase.

\subsubsection{Phase 1} 
We found a significant main effect of the conditions in Phase 1 (${\chi}^2_{(3)} = 9.03$, $p=.028$). 
A post-hoc pairwise comparison with a Mann–Whitney U-Test shows that the coding time ($M=16.06\ mins$) for \condC was significantly faster. In particular, we found a significant difference between \condC and \condA ($M=19\ mins$, $U=59.0$, $p=.005$). We also found a significant difference between \condC and \condB ($M=18.13\ mins, U=51.5, p=.044$). 
No significant differences were found between \condD and other conditions.

Overall, the average coding time for Phase 1 with AI conditions was decreased by 4.6\%-15.5\% compared to the baseline \condA. This also meant that AI conditions (B, C, D) resulted in a 9.0\% faster coding time on average.

\subsubsection{Phase 2}
While we observed that \condD was the fastest in the discussion phase, a Kruskal-Wallis test \textbf{did not show any significant main effect} of the condition on time. The time used for discussion in Phase 2 was between 23.25 mins for \condD and 31.13 mins for \condA (see Figure~\ref{fig:separatecodingtime}).

\subsubsection{Phase 3}
Phase 3 was overall very fast, ranging from 5.63 mins for \condD to 7.19 mins for \condA. We \textbf{did not find any significant main effect} of the conditions on time ($p=.661$).

\subsection{Inter-rater Reliability}

\begin{figure}[htbp]
    \centering
    \includegraphics[scale=0.8]{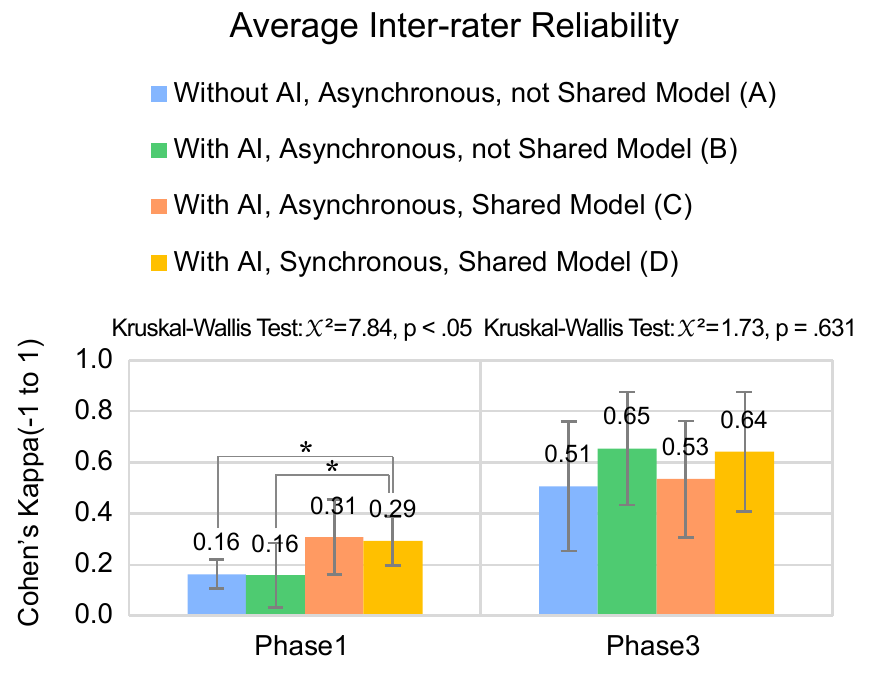}
     \caption{Average Inter-coder Reliability after Phase 1 and after Phase 3. Error bars show .95 confidence intervals. We report the results of the individual Kruskal-Wallis tests, and, if necessary,  pairwise comparisons, where $*: p < .05, **: p < .01$.}
    \label{fig:IRR}
\end{figure}

The average IRR ranges from 0.16 to 0.31 on average in Phase 1. The IRR then increased to 0.51-0.65 by the end of Phase 3, after discussion (Phase 2) (see Figure \ref{fig:IRR}).

\subsubsection{Phase 1} 
A Kruskal-Wallis test shows that there \textbf{is a marginal main effect on the IRR} from the four conditions in Phase 1 (${\chi}^2_{(3)} =7.85$, $p=.049$). Post-hoc pairwise comparisons showed that IRR scores are a bit higher for AI conditions with Shared Model: IRR in \condA was significantly lower than \condD ($U=9.0, p=.015$). IRR in \condB was significantly lower compared to \condD ($U = 13.0, p=.049$). 

\deleted{Although \condC showed a trend towards higher values compared to \condA and \condB, the difference was not statistically significant ($U = 16.0, p=.105$).}

\subsubsection{Phase 3}
We did not observe \textbf{any main effect} between the four conditions for the IRR, ranging from 0.51 in \condA to 0.65 in \condB ($p=.631)$.

\subsection{Code and Subcode Diversity}
\begin{figure}[htbp]
 \centering
 \includegraphics[scale=0.7]{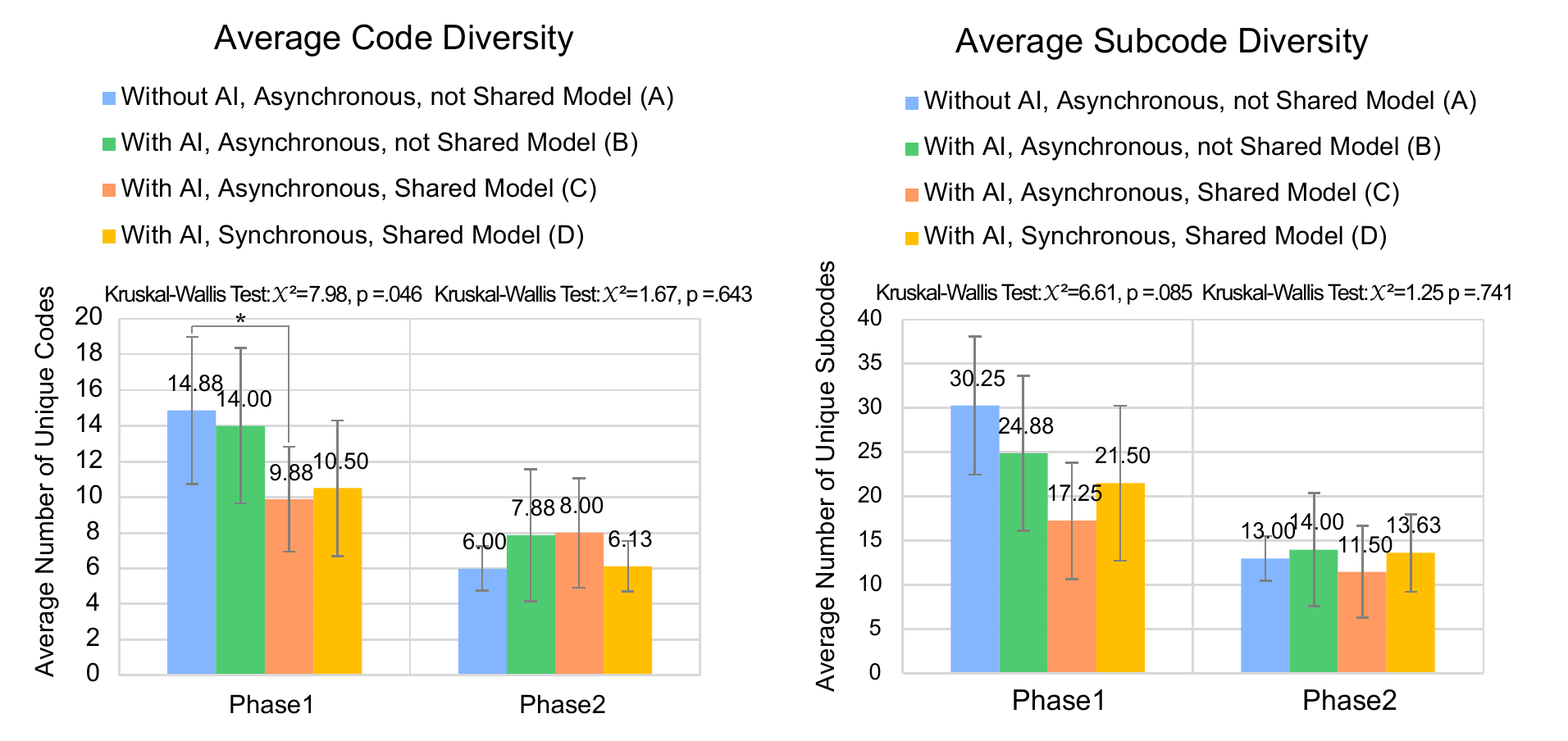}
 \caption{Average Code and Subcode Diversities in Phase 1 and Phase 2. Error bars show .95 confidence intervals. A Kruskal-Wallis test is conducted for the main effect in each phase. Pairwise comparison is performed using Mann–Whitney U-Test with a two-sided alternative, where $*: p < .05, **: p < .01$.}
 \label{fig:diversity}
\end{figure}

The terms, code (i.e., first-level code) and subcode (i.e., second-level code), are as defined as per section \ref{sec:initial_codebook} and \ref{sec:dvs}.
The code and subcode diversity results are summarized in Figure~\ref{fig:diversity}. Our focus will be primarily on Phase 1 and Phase 2. We have opted not to include Phase 3 in our discussion, given that it employs the same codes and subcodes as Phase 2, thus having similar diversity.

\subsubsection{Phase 1}
A Kruskal-Wallis test \textbf{shows a significant main effect} of the condition on code diversity in Phase 1 (${\chi}^2_{(3)}= 7.98$, $p=.046$). Pairwise comparisons show that the observed diversity was lower for conditions with AI and Shared model: the code diversity for \condC ($M=9.88$ unique codes) was significantly lower than for \condA ($M=14.88$, $U=54.0$, $p=.023$).

\deleted{Despite observing a decreasing trend between \condC and \condB ($M=14.00$) as shown by the reduced averages, the results missed the threshold for statistical significance ($U=50.0$ $p=.082$).}

\deleted{Similarly, we observed a decrease in code diversity for \condD ($M=10.50$), but again, this difference was not statistically significant when compared with either \condA ($U=50.0, p=.065$) or \condB ($U=47.5, p=.112$).}

Subcode diversity \textbf{did not seem to be impacted} by our four conditions in Phase 1 (${\chi}^2_{(3)}= 6.61$, $p=.085$). However, it has an average number of unique subcodes ranging from 17.25 in \condC to 30.25 in \condA. 

\subsubsection{Phase 2}
A Kruskal-Wallis test shows that \textbf{there was no main effect in terms of Code Diversity} after Phase 2 (${\chi}^2_{(3)}= 1.67$, $p=.64$). The number of items decreased compared to Phase 1, with the average ranging from 6.00 (\condA) to 8.00 (\condC).
\textbf{This lack of main effect was also visible for subcode diversity} (${\chi}^2_{(3)}= 1.25$, $p=.74$) with the number of subcodes being between 11.5 items (\condC) and 14.00 items (\condB).

\subsection{Code and Subcode Coverage}
\begin{figure}[htbp]
 \centering
 \includegraphics[scale=0.7]{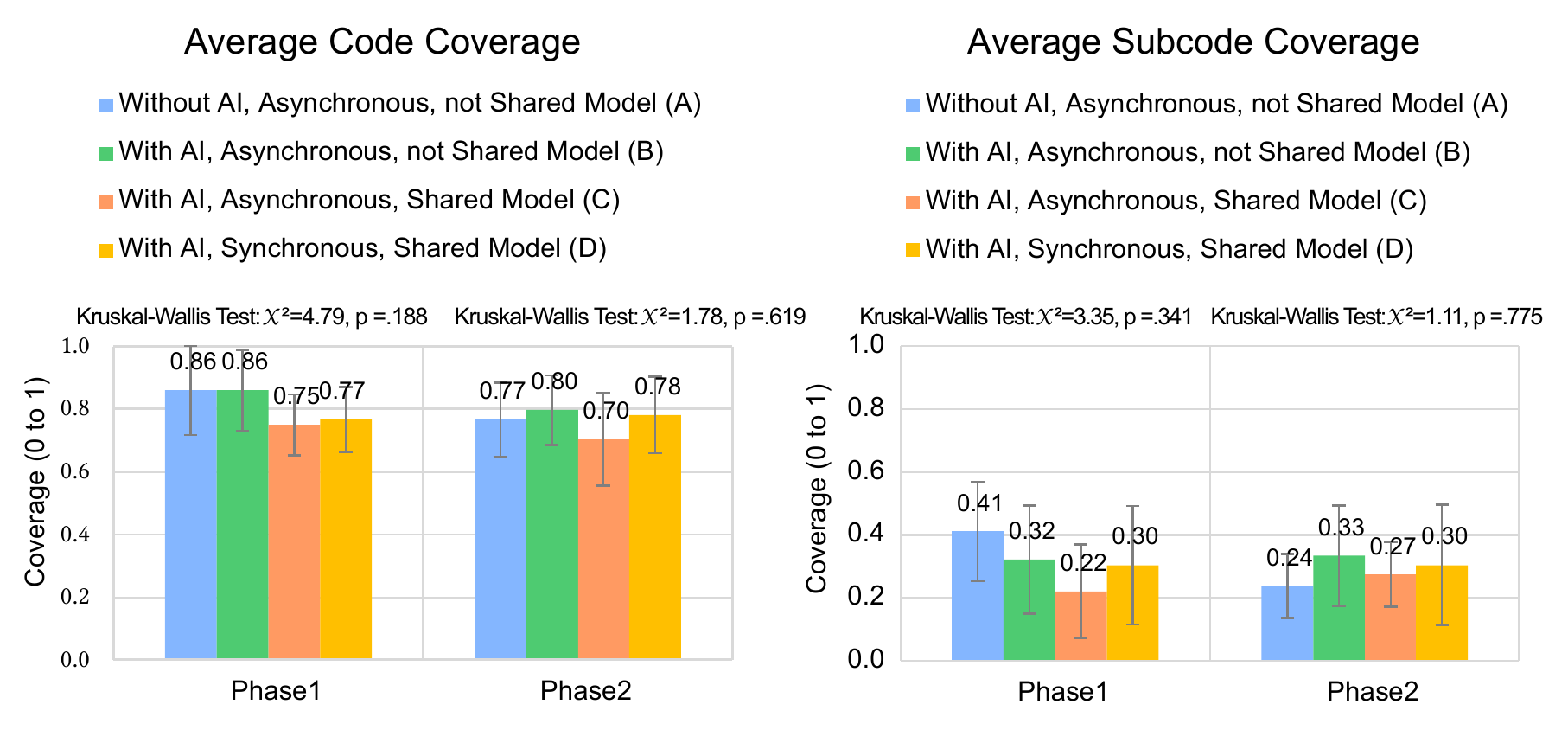}
 \caption{Average Coverage of Code and Subcode in Phase 1 and Phase 2. Error bars show .95 confidence intervals. 
 We report the results of the individual Kruskal-Wallis tests.}
 \label{fig:coverage}
\end{figure}

We report and summarize the average code and subcode coverage in Figure~\ref{fig:coverage}. 

\subsubsection{Phase 1}
We found that \textbf{no main effect for code coverage} between the four conditions in Phase 1 ($\chi^2_{(3)}= 4.79$, $p=.19$), where the coverage of code ranged from 0.75 (\condC) to 0.86 (\condA).

\subsubsection{Phase 2}
Similarly, \textbf{no main effect was found} after Phase 2 (${\chi}^2_{(3)}= 1.78$, $p=.62$) with coverage ranging from 0.70 (\condC) to 0.80 (\condD).

\section{Triangulation with Qualitative Results}

In summary, our quantitative results reveal nuanced disparities among our four conditions with respect to time duration, \irr, and code diversity in Phase 1. However, no significant differences were discerned within \condA and \condB, \textbf{or} \condC and \condD. Most observed variations, therefore, were between \condA and \condB \textbf{on one side}, and \condC and \condD \textbf{on the other}. 

\added{Despite the modest size of the statistically significant differences observed in our study, it is worth noting that even such small results can have meaningful implications \cite{altman1995statistics, hackshaw2008small}. The trends we discerned lead us to a two-fold set of primary findings: First, we found that in the context of our proposed CQA conditions, AI without a shared model may not improve coding efficiency as effectively as the shared model. Second, we found that combining AI with a shared model could potentially accelerate coding speed and achieve a higher level of initial \irr. However, this advantage came with a slight reduction in code diversity during the code development phase. To confirm our primary findings, we employ a triangulation method that combines qualitative results, as discussed in the following.}

\subsection{Lower Initial Coding Time}
While our results indicate that only \condC significantly decreased the coding time in Phase 1, with \condD not reaching statistical significance, there is a suggestive trend that participants under conditions with both AI and a shared model tend to engage in less discussion time in Phase 2 (refer to Figure \ref{fig:separatecodingtime}) and less coding time overall (refer to Figure \ref{fig:totalcodingtime}). The observed decrease ranges from 13.8\% to 18.9\% for total time, 4.6\% to 15.5\% for Phase 1, and 12.8\% to 25.3\% for Phase 2. However, this trend does not extend to Phase 3.
To validate these preliminary observations, a follow-up study with larger datasets would be beneficial in the future.

Additionally, it is important to note that due to study constraints, we had to enforce a time limit of 20 minutes for Phase 1, leading to the discontinuation of the coding task for four pairs who did not complete it within the regulated time (see section \ref{sec:dvs}). Consequently, the differences observed between \condA, which closely adhered to the 20-minute limit in Phase 1 (refer to Figure \ref{fig:totalcodingtime}), and the other conditions may actually be more substantial in real-world usage. However, further data would be required to support this conclusion.

\subsection{Higher Initial IRR}
Likewise, participants using AI \& Shared Model tended to exhibit higher \irr in Phase 1. The increased \irr in \condC and \condD during this phase can be attributed to participants' ability to leverage the shared model, allowing them to update their codes more effectively. This early involvement in the negotiation and merging process, facilitated by the shared model, enables participants to reach an agreement and shared understanding sooner in the initial coding stages. One participant from \condC noted, \textit{"After I wrote (a code), I would check if the suggested codes are better."} (P33 in \condC). Similarly, in \condD, another participant mentioned, \textit{"It could have another word for `introduction'. For example, maybe my partner will say `intro'. But if you want to formalize things, then we realized that introduction is a more formalized code."} (P31 in \condD).

\subsection{Lower Diversity}
While the shared model enabled participants to save time and reach code convergence in the early phases of coding, it also resulted in lower code diversity. On average, in comparison to our baseline condition \condA (Mean = 14.88), the total number of unique codes decreased by 5.00 (in \condC) and 4.33 (in \condD). This reduction can be attributed to the higher overall agreement among participants, which naturally limits the variety of codes used. Additionally, the usage of suggested codes, shared among participants in \condC and \condD, further contributes to the decrease in code diversity.

\subsection{Effect of Synchrony}
It is interesting to note that the synchrony of coding (Synchronous or Asynchronous) did not seem to have an impact on CQA performance, as we did not observe any significant differences between \condC and \condD. Nevertheless, we did observe effects from the qualitative results.

First, we noticed that some coders in \condC tended to rely more on the AI code suggestions, rather than proposing their own codes. This behavior might be attributed to their reliance on a model extensively trained using the first coder's codes. As a result, participants expressed concerns about potential bias introduced by the shared model and expressed worry that the code diversity and coverage in the open coding process (Phase 1) would be reduced.
\textit{"Bias was a bit problematic. Because when you're given suggestions that you might not need, but it shows it has a high confidence level, then subconsciously, I guess you might try to incorporate it [...] which you might not have done otherwise."} (P48 in \condC). This concern is substantiated by our quantitative results.

Second, differences in coding speed could contribute to a similar effect, as slower coders may end up reusing code generated by faster, synchronous coders. As one participant explained, \textit{"We both generated a common knowledge base... but because I was doing slower, then a suggestion coming out with a 0.9 confidence score, which is the code I would have written. That’s why I feel biased."} (P30 in \condD).

It is crucial to note that the coders were not aware that the suggestions originated from the first coder's input, believing them to be AI-generated until it was revealed in the post-interview. We believe that leveraging the shared model is valuable; however, if the second coder was made aware of whether a suggestion originated from the AI or a previous coder, could alter their level of reliance on the suggestions and potentially impact the dependency between both coders.

\subsection{\added{Positive feedback from Shared Model Conditions}}
\added{While the use of a Shared AI model may involve a trade-off in terms of coding speed, \irr, and code diversity, it is notable that the conditions utilizing Shared AI models resulted in more positive experiences among participants.} In both \condC and \condD, 5 out of 8 pairs reported a smooth and swift coding process with the system.

\added{Participants elaborated on the reasons behind their positive experiences.}
For instance, the shared model streamlined the coding process by enabling participants to reuse their own or their partners' previous codes. As one participant noted, \textit{"Because it already has the options that I have entered before, it's faster if I want to add the same code in other paragraphs. I don't have to keep referring to what I wrote above."} (P37 in \condC). They expressed appreciation for the shared model's ability to offer timely suggestions, which substantially assisted them in refining their code expressions. This sentiment was aptly encapsulated by one participant who remarked, \textit{"It sometimes has a better phrase or better word. You can just take from that."} (P34 in \condD). 

The shared model also facilitated coders in aligning their understanding with their partner's during the coding process. As one participant explained, \textit{"For me, yes, coding efficiency was improved. Because my partner was coding just the main points... When I was doing my own coding, I could see [these] main points. It definitely helped me understand what was going on." }(P17 in \condD). This can also prompt them to compare their own viewpoints against a broader range of perspectives. P48 in \condC noted, \textit{"So [AI] suggested \texttt{weakness} [as a code], I thought it could be \texttt{society}. I want to say this kind of inconsistency is just a difference of opinion [of what] terms or labels here." }

In contrast, the majority of participants expressed a neutral or negative attitude towards coding efficiency, while only one participant (P63) from \condB perceived an improvement.

\subsection{Similarity of Codebooks across Conditions}
Overall, our main DVs allow us to understand how AI might help with coding. However, it can still be hard to find out whether the results across our four conditions are similar or at least comparable.
We thus identify the 7 most common first-level codes in the formed codebook for each condition (see Appendix Table~\ref{table:coderanking}). 
Overall, the themes are rather similar (despite a slightly different ranking), suggesting that the results are consistent across the four conditions. Most code categories appeared in all four codebooks, e.g. introduction, leadership, weakness, hiring.

\section{Discussion}
\label{sec:discussion}

\subsection{Trade-off: Coding Efficiency vs. Coding Quality}
Overall, the introduction of AI with Shared Model has the potential to streamline the early coding process, saving time and fostering early consensus on codes, as reflected by higher initial IRR. However, this comes at the cost of reduced initial code diversity. We recognize the inherent trade-off between Coding Time \& IRR and Diversity — essentially, the balance between \textbf{coding efficiency and coding quality} — as a persistent characteristic of the early stages of CQA, particularly when seeking shared AI mediator assistance.

\subsubsection{AI \& Shared Model Fosters Strong Discussions}
Previously, we referenced Zade et al., who identified two types of disagreements that can occur during coding: `diversity' (varying interpretations of a single core idea) and     `divergence' (distinct core ideas) \cite{zade2018conceptualizing}. 

For the former, we found users often favor consistency in coding outcomes across coders over excessive diversity. In this context, the shared model proves beneficial, aiding coders in refining their phrasing.
For the latter, we noted that while coders appreciated AI code suggestions—rooted in their collective coding history—they still actively formulated their own codes, especially when facing divergent disagreements.

This process encourages real-time comparison and validation, prompting coders to critically reassess their coding decisions against alternate perspectives.
When incorporated into subsequent discussions, these reflective insights enrich the overall dialogue. As a result, AI can serve a crucial role as a facilitator in human-to-human collaboration, stimulating more substantive and engaging discussions.

\subsubsection{Potential Pitfalls}
While AI and Shared Model bring considerable benefits, it is crucial not to overlook the following two potential challenges.

\paragraph{Reduced Diversity.} We argue that code diversity, by introducing varied perspectives, plays a crucial role in enhancing coding quality within the CQA process \cite{richards2018practical, anderson2016all}. It is crucial to ensure that coding practices do not excessively compromise coding quality or diversity. Otherwise, users may resort to traditional tools that ensure coding quality through extensive discussions between collaborators on a line-by-line and code-by-code basis. Therefore, we call for the need for further research in this area to address this challenge. For instance, one approach is to enable the system to generate synonyms or keywords of potential codes as references for coders, apart from the original code suggestions \cite{marathe2018semi}. Moreover, allowing the system to easily highlight nuanced differences and compare text selections, by linking back to prior data with simple clicks, can deepen coders' understanding and thus yield more insightful codes.

\paragraph{Over-reliance.} Moreover, we must acknowledge the potential risk of users becoming overly dependent on the system. This tendency can be exacerbated when users aim to increase their speed, encounter challenges in formulating code names, or face discrepancies in coding speed among coders. Such over-reliance could potentially decrease the propensity to seek different opinions, leading users to only choose from system suggestions, particularly during asynchronous coding. This also raises concerns about a potential loss of diversity and nuance in coding, especially in loosely-defined coding tasks where multiple iterations may be necessary to reach consensus on a codebook. In future studies, measures to curb over-reliance on the AI system could include providing explanations for code suggestions \cite{vasconcelos2023explanations}, interface warnings to indicate excessive system use, or even temporary system disabling if over-reliance is detected.

\subsection{Is AI \& Shared Model the best for CQA? Considering Different Application Scenarios}

In light of the points discussed in the previous subsections, we do not argue that AI \& Shared Model is the best solution for CQA. 

\subsubsection{Supporting Different Contexts with Different Independence Level.}
Our evaluation revealed that the four collaboration methods effectively simulated four distinct levels of independence, each potentially desirable in different real-life scenarios (see section \ref{sec:interview}). In conditions without AI or Shared Model (Condition A, B), coders maintained the highest level of independence, with no communication during the code development phase. Conversely, under conditions with AI \& Shared Model (Condition C, D), coders demonstrated a lower level of independence—indirectly connected and communicating through the AI mediator.

The evaluation of different levels of independence yielded varying results. A higher level of independence was found to generate greater code diversity with lower efficiency, while a lower level of independence resulted in lower code diversity with higher code efficiency. 
Align with this, the feedback obtained from participants during the interview (see section \ref{sec:predefined}) indicates that they usually compromise some independence in favor of enhanced coding efficiency when time constraints are present.

The observation, coupled with our formative interviews, has enabled us to discern two distinct scenarios in the context of CQA: efficiency-oriented and creativity-oriented. 

In the former, researchers facing time constraints or requiring both qualitative and quantitative results may be inclined to sacrifice some rigidity in coding for improved coding efficiency. As a result, they might prefer alternative methods, such as using pre-defined themes in analysis, over strict adherence to the Grounded Theory process, which is often seen as more inductive, requiring meticulous, line-by-line open coding \cite{maguire2017doing}. Therefore, when integrating AI to facilitate human-to-human collaboration, some level of communication between coders may be beneficial and acceptable. However, efficiency should not be prioritized to the point of jeopardizing code diversity or undermining the primary goal of CQA—to glean multiple perspectives.

For the latter, in research domains that might heavily rely on creativity, perspective, and critical thinking (e.g., Psychology, Anthropology), it is crucial to minimize external influences and maintain a high degree of independent thought and self-autonomy. For instance, researchers may choose to use separate coding models that generate code suggestions based on each coder's individual coding history.

In summary, we underscore the need to \textbf{consider varying levels of independence across diverse contexts}, as this can impact the trade-off and balance among different coding outcomes when leveraging an AI mediator to facilitate human-to-human collaboration within the realm of CQA.

\subsubsection{Support Different User Groups with AI \& Shared Model}
We believe that \name has the potential to be beneficial for various user groups. In our study, we involved participants who had limited experience with qualitative coding. Based on the feedback from interviewees, this particular user group frequently engages in CQA analysis and codebook creation (see section \ref{sec:difficulty_slow}).

\paragraph{\name for the Learning purpose of Novices}
During our study, participants expressed a positive reception towards the code suggestions, particularly in the initial stages of the coding process. One participant mentioned, \textit{"Firstly, I did not know what to put because I don't know what is required"} (P49 in \condC). Another participant stated, \textit{"Quite useful, because at least I can start. That's how I'm supposed to do it"} (P36 in \condC). These responses indicate that participants found the system helpful in overcoming the initial challenges of understanding coding requirements and getting started with the analysis.

Moreover, participants reported that the system was easy to use. As one participant remarked, \textit{"It seems pretty easy to use"} (P63). By providing a user-friendly interface, our system addresses the learning curve issues highlighted in prior research \cite{jiang2021supporting, yan2014semi} associated with conventional qualitative QA software like nViVo and Atlas.ti. This suggests potential to mitigate users' reliance on traditional collaboration software like Google Docs and Sheets, which are not specifically designed for qualitative analysis.

\paragraph{\name for Expert Users}
We contend that expert coders might derive benefits from our system, as it has the potential to save their time and facilitate higher levels of agreement through AI mediation. We also speculate that expert users may not rely on the system heavily, thereby mitigating the risk of over-reliance. However, further evaluation with expert users is necessary to substantiate this hypothesis.

\section{Design Implication}
\added{Beyond insights into human-to-human collaboration via AI mediation, we also identify design implications for human-AI collaboration, such as the impact of coding granularity and the cultivation of trust between humans and AI.}

\subsection{\added{Impact of Coding Granularity on Human-AI collaboration} \deleted{Unit of Analysis and Code Level}}
Coding granularity includes the unit-of-analysis (UoA) and code specificity.
The UoA delineates the level at which text annotations are made, for example, on a flexible or sentence level \cite{rietz2021cody}. Code specificity refers to the varying degrees of detail in a code, for instance, it can range from a broad code to a more specific one.
In our evaluation, we did not regulate the UoA and code specificity and left the selection of text and codes open. This approach mirrors real-world coding processes, where individuals often apply codes at various units and code specificity~\cite{rietz2021cody}. 
We noted two significant implications arising from this setup.

\subsubsection{\added{Establishing Optimal Coding Granularity for both AI and Human Coders}\deleted{Extreme Codes Affect Performance of AI}}
\added{We first observed a significant variation in coders' interpretation and application of codes, from broad generalities, which often lack informational depth, to more specific interpretations that may not apply universally.} A participant clarified, 
\textit{"So let's say for `leadership' code, right? We should write like `introduction to leadership' and then it can be applied across the 200 (transcripts). If you write like `introduction to event planning', we can't use it, as not everybody organized [event]."} (P17 in \condD).

\added{AI generates suggestions by learning patterns and structures in the data it has been trained on. They are more likely to generalize based on the most prominent features or patterns in the data instead of 
understanding context, nuance, and the complexity of human language and emotions, as noted by a participant in Jiang et al.'s study \cite{jiang2021supporting}. This limitation can have significant implications for how coders approach their work.
If coders become aware that AI systems are more likely to suggest or recognize broad codes, they might deliberately write broader codes to ensure compatibility with AI suggestions, which could lead to oversimplification of the data and possibly missing out on nuanced or intricate details. It also highlights the necessity for human involvement in AI-assisted qualitative coding. The human coder's role becomes critical in guiding the AI, providing the nuance, context, and deeper understanding that the AI may lack. Strategic considerations include the level of specificity in code selection and the possibility of optimizing codes that can be efficiently processed by both AI and human understanding. The goal is to design an AI-assisted system that is not completely dependent on AI but instead integrates the strengths of both AI and human coders to overcome the identified limitations.} 

\subsubsection{\added{Impact of Coding Granularity on IRR Calculation}}
Another challenge we encountered was the evaluation of quality. This issue becomes particularly problematic when calculating IRR, as precise numeric codes on similar levels of the text are usually required for its calculations. When coders choose codes or text of varying levels, it can lead to ambiguity \cite{ceccato2004ambiguity} as it becomes challenging to determine if they've assigned identical codes to a particular unit. \added{To address this challenge, a potential strategy is to regulate coding units or pre-discuss "soft rules" for agreed-upon levels of units (e.g., sentences, paragraphs) before conducting coding \cite{kurasaki2000intercoder, o2020intercoder}. However, to maintain user flexibility in our study, we opted to map the assigned codes to sentence level after coding for IRR calculation.  Future research can explore the combination of these two methods to maximize the advantages they bring.}

\subsubsection{\added{Impact of Coding Granularity on Stability of Suggestions}}
\added{We noted occasional system instability due to frequent retraining, which could change suggestion order or composition and disrupt established interaction patterns.} One participant noted, \textit{"When I selected one sentence of this paragraph, the code is there. But when I select the whole paragraph, the [same] code is not there [any more]."} (P47 in \condC). 
\added{This instability presents challenges for coders' user experience who predictively interact with the system and rely on prediction stability \cite{liu2022model}. However, users primarily using the system for decision-making aid and consistently choosing from the suggestion list might have minimal reliance on prediction stability. For them, minor alterations might not pose significant issues, as long as the system continues to provide relevant and accurate suggestions. Furthermore, a user's dependence on prediction stability could also evolve with familiarity with the system. Novice users might rely heavily on the system's suggestions stability, while experienced users might develop their own interaction strategies.}
\added{From the model's perspective, prediction stability depends on the extent of data changes between training iterations. If the newly added data closely mirrors the pre-existing data, the model's predictions could remain stable. Conversely, if the new data significantly diverges, predictions might undergo noticeable changes. Employing strategies like incremental learning, where the model learns from new data without forgetting previous knowledge, could potentially maintain stable predictions \cite{giraud2000note}.}

\subsection{\added{Trust and User Expectations}\deleted{Managing Users' Expectation}}

\added{While we strive for maximum accuracy in AI systems, it is important to acknowledge that achieving perfection, especially in tasks involving subjective data, is challenging \cite{chen2018using, jiang2021supporting}.}

\subsubsection{Calibrate Users' Expectation Before Coding}
Previous research has demonstrated that unrealistic user expectations can lead to reduced user satisfaction with AI systems \cite{kocielnik2019will, grimes2021mental, ashktorab2020human, cheng2019explaining}.
In our evaluation, participants exhibited a notable initial expectation regarding the AI's capability to provide suggestions. If the system failed to meet their expectations, participants resorted to manually entering their own codes. Additionally, the occasional inaccuracies in the AI's suggestions brought confusion and might damage their confidence in the AI. \textit{"For one sentence I thought it was "interest". The number [for the other suggestion] was like 0.9 but for "interest" was like 0.00 something. I was a bit confused."} (P62 in \condB). 

Additionally, participants expressed a desire for more timely results when requesting code suggestions. However, our current "retrieve-train-predict" process takes at least 10 more  seconds, even when utilizing a GPU. This delay in providing suggestions may have impacted participants' willingness to utilize the AI suggestions.
\textit{"One difficulty is the code I put in takes a while to come out when I want to use it again for another passage."} (P38 in \condC).

In future work, in addition to enhancing the accuracy and stability of the system, managing users' expectations will be crucial. One approach could involve calibrating users' expectations by providing information about the system's capabilities, expected accuracy, and the possibility of errors. Furthermore, making the suggestions more explainable and interpretable could provide users with insights into the underlying reasoning, potentially easing users' doubts, distrust, and frustration \cite{knowles2015models, golafshani2003understanding, lubars2019ask, liao2020questioning}.

\subsubsection{Can Imperfect Suggestions Help?}
Research suggests that even imperfect AI can still provide valuable assistance to users \cite{kocielnik2019will}. 
In our case, when AI suggestions are in conflict with the assumptions made by coders, it indicates the possibility of either partially incorrect or completely incorrect suggestions. When AI suggestions partially align with users' thoughts, they can select the suggested codes and implement minor adjustments. Therefore, these suggestions can still assist users in making code decisions, and this valuable user input can contribute to improving the model's performance in subsequent training.
When AI suggestions are completely not matched, users have the option to bypass the suggestions and manually enter their own codes.

However, some coders may not be aware of the aforementioned strategy. When they encounter imperfect suggestions, their trust in the AI system can diminish.
\textit{"I guess I did not use [the AI suggestions] much, because the words I needed wouldn't not be suggested. But I guess if one suggestion is a bit close to what you are thinking, it’s enough. But if you want to put your exact thoughts, then I guess doing it manually would be better."} (P48 in \condC).
Sometimes, a relevant code suggestion may not be among the top five suggestions but could instead appear within the top ten. Therefore, it would be beneficial for users to have access to a longer list of codes upon request. While this feature may not be frequently utilized, it would grant users more control over the system.
\textit{"I think for myself, [the system] wasn't that helpful... because there's only five in the drop-down list. Even if I want something that is exactly the same words, but it's not in the top five recommended, I cannot get it. There isn't a scroll down or something."} (P32 in \condC).

Overall, it is important to communicate to coders about the capabilities of the AI system and how to effectively respond to imperfect code suggestions. Such information can enhance coders' understanding of AI system's capabilities and limitations, allowing them to fully leverage it.
\section{Limitation and Future Work}

\added{This work has limitations. First, our research primarily involved novices participating in the system evaluation. There were several reasons for this choice: 1) based on our initial interviews, training novice users and integrating them into the CQA team is a common practice (see section \ref{sec:difficulty_slow}); 2) in our investigation of collaboration factors within the CQA context (i.e., With/Without AI, Synchrony, Shared Model/No Shared Model), designing a between-subject design was necessary. Ensuring a similar baseline experience among participants was also essential to maintain fairness across conditions. In the study, we took measures to promote accurate task completion: a) we provided thorough CQA training to all participants in each session, aiming to equip them with the skills necessary to efficiently carry out the tasks; b) we monitored the participants' coding results and the quality of their work closely during the task to minimize disruption to their coding process (see section \ref{sec:study_design:procedure}); c) after collecting data, we carried out quality checks to verify the reliability and completeness of the results (see section \ref{sec:study_design:data_analysis}).}

\added{However, despite these measures, it's important to acknowledge that our findings might differ from those obtained from seasoned CQA experts. For example, in some isolated cases, we observed a slight over-reliance on the system by novices. However, even when employing novices, the need for discussion and collaboration persists. Therefore, experts could also derive benefits from the time-saving capabilities of our system.
Future research should broaden the user base to include expert-expert and expert-novice pairs, as well as native-nonnative speakers, to explore other potential advantages of the various collaboration diagrams. }

\added{Another limitation of our study is the lack of substantial statistical significance, which can potentially be traced back to multiple factors. 
One key aspect is the inherent complexity of human collaboration in the context of CQA tasks, which is multifaceted and nuanced, carries considerable significance. For example, the diversity of reactions among coders to AI-suggested codes, or the differences in the pace at which individual coders learn and adapt to the coding process. Additionally, coders might have widely different coding strategies that influence their coding speed and the quality of their output. Furthermore, while we opted for a potentially optimal number of AI suggestions for the \namef design in this iteration, the influence of the number of suggestions on user behaviors is an important design aspect to consider. It could also be interesting to examine the impact of revealing AI suggestions either before or after users select the text. 
}

\added{Another consideration lies in system-related factors, including the frequency of training updates and delay in achieving stable model accuracy, which are certainly significant.  It is worth noting that improving model performance on subjective annotation continues to be a key challenge in fields like NLP~\cite{davani2022dealing} and HCI~\cite{chen2018using}, and this aspect remains under explored in the context of CQA. In future work, it would be advantageous to incorporate data augmentation technologies or other NLP pipelines to enhance model performance. Moreover, the promising potential of advanced LLMs, e.g., GPT-4\footnote{\url{https://openai.com/research/gpt-4}}, which demonstrate exceptional capabilities in understanding and generating text \cite{zhang2023visar}, should not be overlooked. They could be harnessed to facilitate AI-assisted qualitative coding\footnote{\url{Announced on 28th March 2023: https://atlasti.com/ai-coding-powered-by-openai}} as well as CQA \cite{gao2023collabcoder}.}

\added{These individual differences and system-related factors, though subtle, can cumulatively exert a profound impact on collaborative coding dynamics and the detection of the final significant difference. However, they remain largely unexplored. While we recognize their significance, our main emphasis in this work lies in unearthing the possibilities and influence of AI mediation on human collaborative dynamics within a CQA context. We strongly recommend further research in this field, given the significant role of CQA in qualitative research and the current focus primarily on AI's application in individual qualitative analysis. Our goal is to pinpoint and investigate the research gap, rather than establish a definitive or arbitrary methodology for AI-assisted CQA.}
\section{Conclusion}
In this work, we delve into the AI-assisted human-to-human collaboration within the context of CQA and assess various collaboration modes between coders. To the best of our knowledge, this marks the first attempt to investigate the role of AI in the collaboration of qualitative coding, as previous research primarily concentrated on individual coding with AI. 
In pursuit of this goal, we initially gained insights into coders' CQA behaviors, challenges, and potential opportunities through a series of semi-structured interviews. Following this, we developed and implemented a prototype, \name, designed to provide coders code suggestions based on their coding history. We further introduced four collaboration methodologies and evaluated them using a between-subject design with 64 participants (32 pairs). This led to the identification of the trade-off between coding efficiency and coding quality, as well as the relationship between independence level and the coding outcomes under varying CQA scenarios. We also highlighted design implications to inspire future CQA system designs.

\bibliographystyle{ACM-Reference-Format}
\bibliography{paper/reference}

\newpage
\appendix

\newmdenv[linecolor=black,innerlinewidth=0.5mm,roundcorner=4mm,backgroundcolor=gray!10,innertopmargin=\topskip]{mybox}

\section{Study Protocol}
\label{appendix:study_protocol}

\begin{mybox}

\subsection{Welcome to AIQA Study!}

Qualitative analysis (QA), a common method in human-computer interaction and social computing research, involves a key process known as coding. This procedure is crucial for discerning patterns and extracting insights from qualitative data, though it's traditionally labor-intensive and time-consuming. In recent years, researchers have introduced Artificial Intelligence (AI) to enhance the efficiency of this process. However, they've largely overlooked the collaborative aspect of the coding process. Our project seeks to bridge this gap by offering an AI-based tool to streamline collaboration among coders. Utilizing AI to facilitate this interaction could potentially improve coding efficiency, potentially saving considerable time for QA researchers.

\subsection{Task Introduction}

You are a pair of researchers who are trying to perform qualitative analysis on interview transcripts of students undergoing a preparatory mock interview. The research question is to find general qualities of candidates (include good and bad ones).

Your task is to code the sentences so that we may obtain a meaningful analysis of the transcripts. Here's an example. If we were to analyse meeting transcripts, a likely thing to be said during a meeting might be:

\begin{quote}
    "Ok, can Alice please follow up with Bob on the designs".
\end{quote}

\begin{itemize}
    \item A reasonable way to code this sentence could be \textbf{Action Items}.
    \item A succinct subcode or description could be \textbf{"A person was asked to follow up on a task"}.
    \item This sentence would then be added as one of the \textbf{examples} of Action Items code.
\end{itemize}

Participants are also presented with a sample codebook table, specifically Table 1 from DeCuir-Gunby et al.'s work \cite{decuir2011developing}.

\subsection{Introduction to Three Phases}
In the introduction, the instructor presents various strategies for employing \name, which are designed to support distinct conditions across three phases of CQA.

\subsection{Post-Study Interview Questions}
\begin{enumerate}
    \item What challenges have you encountered during the labeling process when working individually?
    \item What difficulties arise when you engage in collaborative labeling?
    \item In your opinion, how effectively does \name manage these collaborative challenges?
    \item What is your reaction when you encounter a code in the code list that you did not personally contribute?
    \item How would you describe your level of confidence when using the AIcoder?
    \item Has \namef assistance proven beneficial in resolving conflicts that arise during the coding process? If so, how?
    \item How frequently do you utilize \name, and what motivates this usage?
    \item Do you perceive that \name enhances your coding efficiency and collaboration? Could you please elaborate?
    \item Have you noticed a speed increase in the coding process after the formation of the codebook? If so, how has this been achieved?
    \item Any other relevant and improvised questions.
\end{enumerate}

\end{mybox}

\section{Codes in Codebooks Generated by Participants}

To facilitate an intuitive comparison of the results across four conditions, the table below presents the ranking of codes in the formed codebook for each condition.

\begin{table}[htbp]
\footnotesize
\caption{The Ranking of Codes in Formed Codebook of Four Conditions. Only codes in the first level were counted. The codes in every cell are different expressions of one core idea labelled in bold.}
\label{table:coderanking}
\scalebox{0.78}{\begin{tabular}{@{}lllll@{}}
\toprule
Ranking & \begin{tabular}[c]{@{}l@{}} Condition A: \\Without AI, \\ Asynchronous, \\ not Shared Model \end{tabular} & \begin{tabular}[c]{@{}l@{}} Condition B: \\With AI, \\ Asynchronous, \\ not Shared Model \end{tabular} & \begin{tabular}[c]{@{}l@{}} Condition C: \\With AI, \\ Asynchronous, \\ Shared Model \end{tabular} & \begin{tabular}[c]{@{}l@{}} Condition D: \\With AI, \\ Synchronous, \\ Shared Model \end{tabular} \\ \midrule
1 & \begin{tabular}[c]{@{}l@{}}\textbf{Leadership}: \\ Leadership skills(3); \\ Leadership; \\ leadership experiences(2); \\ Leadership experience \\ with poor relevance; \\ Work Experience\end{tabular} & 
\begin{tabular}[c]{@{}l@{}}\textbf{Strengths}: \\ Humble; Motivated; \\ Open-minded; Sociable; \\ very focused; Courage; \\ Reflective; Introspection; \\ Rational; Confidence; \\ Good qualities \end{tabular} & \begin{tabular}[c]{@{}l@{}}\textbf{Leadership}: \\ Leadership training; \\ Leadership skills; \\ Leadership Qualities(2); \\ leadership(2); Leadership role; \\ Leadership experience(2)\end{tabular} & \begin{tabular}[c]{@{}l@{}}\textbf{Leadership}: \\ Leadership experience(4); \\ leadership(3); \\ Leadership skills\end{tabular} \\\midrule
2 & \begin{tabular}[c]{@{}l@{}} \textbf{Weakness}: \\ Using their weakness to \\ their advantages; Weakness that \\ is addressed; Weakness(2); \\ Discussion on weakness; \\ Overcoming weakness; \\ Poor example of overcoming \\ weakness; Weakness and \\ overcoming weakness\end{tabular} & \begin{tabular}[c]{@{}l@{}}\textbf{Weakness}:\\ Weakness(6); \\ slightly impulsive; \\ Indecisive; Shy personality; \\ Bad qualities; overcoming; \\ Overcoming weakness; \\ ways to overcome weaknesses\end{tabular} & \begin{tabular}[c]{@{}l@{}}\textbf{Weakness}: \\ Weakness(5); \\ Candidate's weakness; \\ Overcoming weakness; \\ Weakness and how you overcome\end{tabular} & \begin{tabular}[c]{@{}l@{}}\textbf{Weakness}: \\ Weakness(4); \\ personal weaknesses; \\ Weakness and \\overcoming weakness; \\ How to overcome weakness; \\ Sharing weaknesses\end{tabular} \\\midrule
3 & \begin{tabular}[c]{@{}l@{}}\textbf{Hiring}:\\ Key qualities for hiring; \\ Irrelevant reasons for hiring; \\ Self-marketing; \\ Why candidate should be hired; \\ Strengths and reason for hire; \\ Reason to hire\end{tabular} & \begin{tabular}[c]{@{}l@{}}\textbf{Leadership}:\\leadership(2); Leadership training; \\ Group-oriented leadership style; \\ Leadership and teamwork; \\ leadership experience(4); \\ Leadership skills\end{tabular} & \begin{tabular}[c]{@{}l@{}}\textbf{Introduction}:\\Background; \\ Introduction(3); \\ self-introduction(3)\end{tabular} & \begin{tabular}[c]{@{}l@{}}\textbf{Introduction}:\\Introduction(3); \\ Introduction and Interest; \\ Current status;\\ self-introduction\end{tabular} \\\midrule
4 & \begin{tabular}[c]{@{}l@{}}\textbf{Introduction}:\\Personal introduction; \\ Introduction; \\ Education; \\ Introduction of candidate; \\ Interests\end{tabular} & \begin{tabular}[c]{@{}l@{}}\textbf{Introduction}:\\Introduction(3); \\ Background Information\end{tabular} & \begin{tabular}[c]{@{}l@{}}\textbf{Challenges}:\\Challenges faced(3); \\ Challenging Situation; \\ challenge; Problem recognition; \\ Team working challenge; \\ Lack of resources; \\ Language barriers\end{tabular} & \begin{tabular}[c]{@{}l@{}}\textbf{Teamwork}:\\ Teamwork(2); \\ Teamwork experience(2)\end{tabular} \\\midrule
5 & \begin{tabular}[c]{@{}l@{}}\textbf{Teamwork}: \\ Tendency to help/accommodate\\ teammates; \\ Teamwork(2); \\ Team experience\end{tabular} & \begin{tabular}[c]{@{}l@{}}\textbf{Challenges}: \\ Difficulties; Challenges faced; \\ examples of challenges \\working in a team; \\ Teamwork challenges\end{tabular} & \begin{tabular}[c]{@{}l@{}}\textbf{Hiring}:\\Reasons to Hire Candidate; \\ reason to hire; perfecting herself; \\ Reasons that interviewee \\ should get hired; \\ Why should you be hired\end{tabular} & \begin{tabular}[c]{@{}l@{}}\textbf{Hiring}:\\Hiring Quality; \\ Reasons to hire(2); \\ Hiring decision; \\ Strengths and reason \\ for hire\end{tabular} \\\midrule
6 & \begin{tabular}[c]{@{}l@{}}\textbf{Problem solving}:\\Problem solving; \\ Problem Solving Skills \\with poor relevance; \\ Experienced problem solving skills\end{tabular} & \begin{tabular}[c]{@{}l@{}}\textbf{Hiring}:\\Reasons for hiring; reason to hire; \\ reasons for hiring interviewee\end{tabular} & \begin{tabular}[c]{@{}l@{}}\textbf{Problem solving}:\\Overcome challenges(2); \\ Dealing with the challenge; \\ Problem-solving skills; \\ Problem solving(2) \end{tabular} & \begin{tabular}[c]{@{}l@{}}\textbf{Strengths}:\\Sharing personal strengths; \\ Positive attributes; Strengths; \\ Competitive advantage\end{tabular} \\\midrule
7 & \begin{tabular}[c]{@{}l@{}}\textbf{Interest in role}:\\Candidate's Vague Interest in Role; \\ Candidate has Interest in the Role\end{tabular} & \begin{tabular}[c]{@{}l@{}}\textbf{Interest}:\\Keen in health; interest \end{tabular}& \begin{tabular}[c]{@{}l@{}}\textbf{Strengths}:\\Strengths(2); \\ Candidate's strengths; \end{tabular} & \begin{tabular}[c]{@{}l@{}}\textbf{Challenges}:\\ Challenges faced; Challenge; \\ Challenges and actions taken; \\ handling challenges\end{tabular} \\ \bottomrule
\end{tabular}}
\end{table}

\end{document}